%% file: 00_main.tex
\documentclass[sigconf]{acmart}
\AtBeginDocument{%
  \providecommand\BibTeX{{%
    \normalfont B\kern-0.5em{\scshape i\kern-0.25em b}\kern-0.8em\TeX}}}

\usepackage{hyperref}

\usepackage{latexsym}
\usepackage{url}
\usepackage{paralist}
\usepackage{multirow}
\usepackage[most]{tcolorbox}
\usepackage{mathtools}
\usepackage{adjustbox}
\usepackage{subfiles}
\include{commands}
\usepackage{times}

\usepackage{caption, subcaption}
\usepackage{wrapfig}

\AtBeginDocument{%
  \providecommand\BibTeX{{%
    \normalfont B\kern-0.5em{\scshape i\kern-0.25em b}\kern-0.8em\TeX}}}

\setcopyright{acmcopyright}\acmConference[CHI '22]{CHI Conference on Human Factors in Computing Systems}{April 29-May 5, 2022}{New Orleans, LA, USA}
\acmBooktitle{CHI Conference on Human Factors in Computing Systems (CHI '22), April 29-May 5, 2022, New Orleans, LA, USA}
\acmPrice{}
\acmDOI{}
\acmISBN{}
\begin{document}

%%
%% The "title" command has an optional parameter,
%% allowing the author to define a "short title" to be used in page headers.

\title{Scaling Creative Inspiration with Fine-Grained Functional Aspects of Ideas}

\author{Tom Hope}
\email{tomh@allenai.org}
\affiliation{
\country{}
  \institution{Allen Institute for AI \\ The University of Washington}
}
\author{Ronen Tamari}
\affiliation{
\country{}
  \institution{Hebrew University of Jerusalem}
}

\author{Hyeonsu Kang}
\affiliation{
\country{}
  \institution{Carnegie Mellon University}
}

\author{Daniel Hershcovich}
\affiliation{
\country{}
  \institution{University of Copenhagen, Denmark}
}
\author{Joel Chan}
\affiliation{
\country{}
  \institution{University of Maryland}
}
\author{Aniket Kittur}
\affiliation{
\country{}
  \institution{Carnegie Mellon University}
}
\author{Dafna Shahaf}
\email{dshahaf@cs.huji.ac.il}
\affiliation{
\country{}
  \institution{Hebrew University of Jerusalem}
}

%%
%% By default, the full list of authors will be used in the page
%% headers. Often, this list is too long, and will overlap
%% other information printed in the page headers. This command allows
%% the author to define a more concise list
%% of authors' names for this purpose.
\renewcommand{\shortauthors}{Hope et al.}

%%
%% The abstract is a short summary of the work to be presented in the
%% article.

\begin{abstract}

Large repositories of products, patents and scientific papers offer an opportunity for building systems that scour millions of ideas and help users discover inspirations. However, idea descriptions are typically in the form of unstructured text, lacking key structure that is required for supporting creative innovation interactions. Prior work has explored idea representations that were either limited in expressivity, required significant manual effort from users, or dependent on curated knowledge bases with poor coverage. We explore a novel representation that automatically breaks up products into fine-grained functional aspects capturing the purposes and mechanisms of ideas, and use it to support important creative innovation interactions: functional search for ideas, and exploration of the design space around a focal problem by viewing related problem perspectives pooled from across many products. In user studies, our approach boosts the quality of creative search and inspirations, substantially outperforming strong baselines by 50-60\%.

\end{abstract}

%%
%% The code below is generated by the tool at http://dl.acm.org/ccs.cfm.
%% Please copy and paste the code instead of the example below.
%%
% \begin{CCSXML}
% <ccs2012>
%  <concept>
%   <concept_id>10010520.10010553.10010562</concept_id>
%   <concept_desc>Computer systems organization~Embedded systems</concept_desc>
%   <concept_significance>500</concept_significance>
%  </concept>
% </ccs2012>
% \end{CCSXML}

% \ccsdesc[500]{Computer systems organization~Embedded systems}

%%
%% Keywords. The author(s) should pick words that accurately describe
%% the work being presented. Separate the keywords with commas.
% \keywords{datasets, neural networks, gaze detection, text tagging}
\maketitle

\section{Introduction}
\label{intro-sec}
\input{01_finalintro}

\section{Related Work: Ideation Systems and Cognitive Mechanisms}
\input{02_related}

\label{related-sec}
%\section{Learning fine-grained purposes and mechanisms}
%\label{fine-sec}
%\input{fine_grain_motivation.tex}
%\input{fine_grain_data.tex}
\section{Learning a Fine-Grained Functional Representation}
\label{formmodel-sec}

\input{02_general_formulation}

\subsection{Extracting Spans}
%\tnote{calling this model may again risk the ire of pesky reviewers... Information extraction baselines?}
\label{eval-sec}
%tnote{calling this model may again risk the ire of pesky reviewers... accuracy evaluation?}
\input{03_acc_eval}

\label{search-sec}
\input{04_Search}

\input{05_search_eval}

%\input{motivation_sec.tex}

\section{Exploring the Design Space with a Functional Concept Graph}
\input{06_Hier}
\label{hier-sec}

\input{07_hier_eval}

%\section{Related Work}
%\label{related-sec}
%\input{related.tex}

%\section{Related work}
%\label{related-sec}
%\input{related.tex}

\section{Discussion and Conclusion}
\label{discussion-sec}
\input{08_Discussion}

{%\small
\bibliographystyle{abbrv}
\bibliography{sample-base}
}
\newpage
\appendix
%\section{Reproducibility}\label{sec:appendix}
%\input{Implementation_details.tex}

%\appendix
\section{Technical Appendix}\label{sec:appendix}
\input{09_Supp}

\end{document}

%% file: commands.tex
\newcommand{\remove}[1]{}

\newcommand{\xhdr}[1]{\vspace{1mm}\noindent{{\bf #1.}}} 
\newcommand{\R}{{\mathcal R}}
\newcommand{\Rbb}{{\mathbb R}}
\newcommand{\Fone}{\text{F}_{1}}
 

%% file: 01_finalintro.tex
%, helping human problem solvers identify inspirations and solutions across domains. 

Human creativity often relies on detecting \emph{structural matches} across distant ideas and \emph{adapting} them by transferring mechanisms from one domain to another \cite{gentner1997structure,gentner_relational_2005,chan2011benefits,chan2016analogy}. For example, microwave ovens were discovered by {\it repurposing} radar technology developed during World War II. Teflon, today chiefly used in non-stick cookware, was first used in armament development. Recognizing the potential of this innovation process, major organizations such as NASA and Procter \& Gamble actively engage in searching for opportunities to adapt existing technologies for new markets \cite{pg}. 

Online repositories of millions of products, scientific papers, and patents present an opportunity to augment and scale this core process of innovation. The large scale and diversity of these repositories is important, because inspiration can be found in unexpected places -- for example, a car mechanic recently invented a simple device to ease childbirths by adapting a trick for extracting a cork stuck in a wine bottle, which he discovered in a YouTube video \cite{venema_car_2013}.

However, the predominant way human problem-solvers currently interact with these repositories --- via standard search engines --- does not tap into their potential for augmenting and scaling human ingenuity. Core to this limitation is the representation of ideas in the form of unstructured textual descriptions. This representation hinders creative innovation interactions that require traversing multiple levels of granularity and abstraction around a focal problem, to ``break out'' of fixation on the details of a specific problem by
exploring the design space and viewing novel perspectives on problems and solutions \cite{linseyWordTreesMethodDesignbyanalogy2008,dorstCoreDesignThinking2011,chanBestDesignIdeas2015,kittur2019scaling}.

Toward addressing this challenge, our vision in this paper is to develop a novel representation of ideas that can support \textbf{exploration and abstraction of fine-grained functional aspects in large-scale idea repositories} -- aspects such as the {purposes and mechanisms} of products. More specifically, our goal is to obtain a representation having two key capabilities: {(I)} The representation would be able to automatically disentangle raw descriptions into {\bf fine-grained functional ``units''} that support search and discovery of products that match on certain functions but not others. {(II)} This representation should also allow  navigating the landscape of ideas at {\bf different resolutions} --- enabling users to ``zoom'' in and out at desired levels of abstraction of a given problem and connect to inspirations in seemingly distant areas. 

\begin{figure*}[t]
    %\centering
    \includegraphics[width=0.85\linewidth]{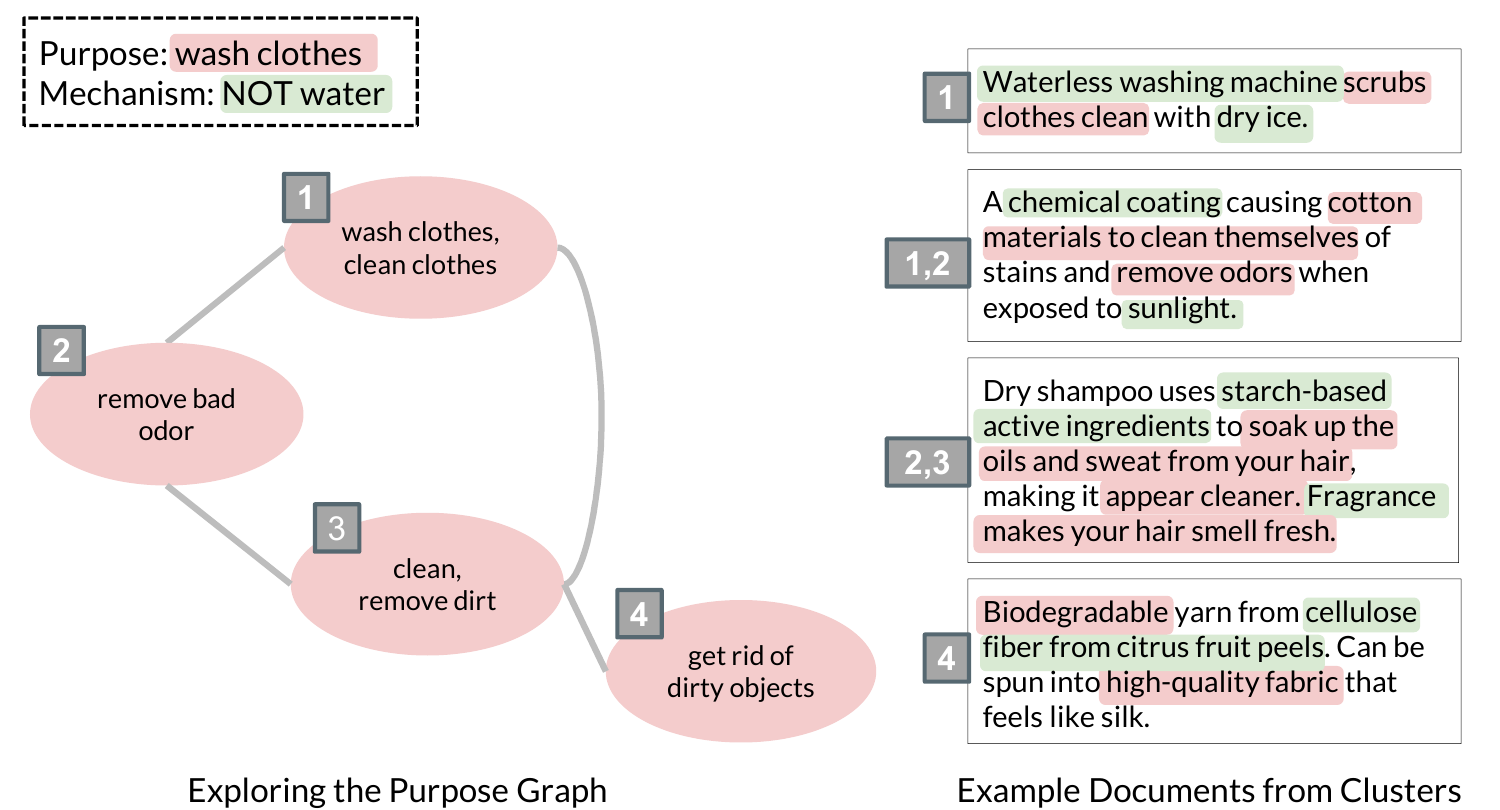}
    \caption{Extracting fine-grained purposes and mechanisms at scale enables mapping the landscape of ideas. Suppose an inventor is looking for a way to wash clothes without water. On the left we see a snippet from the graph of purposes. Each node in the graph represents a cluster of similar purpose spans extracted from documents (labels are manually generated for illustration purposes). Edges reflect abstract structural similarity, capturing co-occurrence patterns of spans in the corpus (see Section \ref{hier-sec} and Figure \ref{fig:medgraph}). On the right we see example documents containing purposes from the four clusters (corresponding cluster numbers appear in boxes). Purpose/mechanism spans in documents are shown in pink/green, respectively. One could find direct matches to the query -- i.e., documents with purpose from cluster $1$ and mechanism not ``water'' (e.g., waterless washing machine using dry ice), or explore neighboring purpose clusters, reformulating the problem as removing odor, removing dirt, or getting rid of the dirty clothes, each resulting in a different set of inspirations.}
    \label{fig:system}
\end{figure*}

As an example, consider an inventor looking for a way to wash clothes without water (e.g., in space, or where water is scarce). Figure \ref{fig:system} illustrates our vision. 
Breaking down product descriptions into fine-grained functions ({\it capability I} above) could allow an automated system to find ideas that match on certain purpose aspects (\emph{washing clothes}) but not certain mechanism aspects (\emph{usage of water}). This could lead to solutions like %dry shampoo, that removes odors, and 
cleaning mechanisms based on dry ice or chemical coating.

Zooming out and abstracting the problem to a more general framing ({\it capability II}) might lead to broader ideas for the problem of \emph{cleaning} such as techniques for removing dirt or odor -- each resulting in novel problem perspectives and inspirations.
In Figure \ref{fig:system}, each node represents a cluster of documents with a similar purpose and the user can explore neighboring clusters to find inspirations (e.g., dry shampoo). This can also expand the innovator's conception of the problem space itself, such as the assumption that clothes should be cleaned and reworn (vs. biodegradable). 

%ining connections between  of functions that maps

In this paper, we develop a scalable computational model of ideas that brings us closer to this vision. We train a neural network to extract spans of text describing purposes and mechanisms in product texts, and use them to build a span embedding representation that allows aspect-based matching between ideas. We then use this representation of \textit{individual} ideas to automatically mine connections between problems and solutions across \textit{entire repositories} and build a ``functional network'' that resembles functional ontologies used in engineering and design ideation \cite{hirtz2002functional,gerickeIntegratedFunctionModeling2017}, which are typically hand-crafted and limited in scale.

Our approach could facilitate many applications in creative innovation due to its ability to decompose idea texts into fine-grained functional aspects, and to surface related problem perspectives at multiple levels of abstraction --- two fundamental drivers of creativity support. In this paper we instantiate the approach in two prototype systems, probing its value regarding each of these capabilities:
\begin{itemize}
\item {\bf Functional aspect-based search for alternative uses.} 
One important use of our novel representation is to enhance the expressivity of search engines over idea repositories.
%For example, consider the inventor looking for a way to wash clothes without water (Figure \ref{fig:system}). Using our system, they could pose a query looking for products with a purpose similar to ``wash clothes'' and a mechanism which is ``NOT water''. This could lead to novel inspirations like dry shampoo, chemical coating that removes odors, and cleaning mechanisms based on dry ice.
This way, our representation could support expressive search for \textit{alternative, atypical uses of products} to identify potentially high-value adaptation opportunities. %similar to the invention of non-stick cookware by repurposing Teflon.

%Our approach allows formulating expressive queries, such as searching for diverse mechanisms achieving a specific purpose.
\item {\bf Exploring alternative levels of problem perspectives.} 
Recent work \cite{gilon2018focab} showed that problem-solvers are often interested in exploring different reformulations of the problem when searching for inspiration.
%(e.g., going from ``compact to ship'' to ``efficient to ship'', ``easy to move'', or ``miniaturization''). 
Our fine-grained span representations facilitate mining of recurring functional relations, such as purposes that are often mentioned together or mechanisms associated with purposes. This level of detail enables us to map the landscape of ideas with a network of purposes and mechanisms, allowing us to automatically traverse neighboring problems and solutions around a focal problem and surface novel inspirations to users. \end{itemize}

Previous work highlighted the importance of functional representations for supporting ideation \cite{linseyWordTreesMethodDesignbyanalogy2008,yu2016distributed, hope2017accelerating, chan2018solvent,gilon2018focab, kittur2019scaling}, but these methods require significant \emph{manual effort} from the user, rely on resources with limited coverage, or have limited expressivity (we discuss this work in more detail in §\ref{related-sec}). 
We seek to advance the state of the art by developing a novel representation that is both expressive and scalable, and exploring the applications it unlocks. We believe our representation may serve as a useful building block for novel creativity support tools that can help users find and recombine the inspirations latent in unstructured idea repositories at a scale previously impossible.

\textbf{In summary, in this work we contribute}:
\begin{compactitem}
\item A novel computational representation of ideas with fine-grained functional aspects for purposes and mechanisms.

\item Empirical demonstrations of the flexibility and utility of the representation for 
computational support of core creative tasks: (1) searching for \textit{alternative, atypical product uses} for potential adaptation opportunities; and (2) creating a \emph{functional concept graph} that enables innovators to explore the design space around a focal problem. Through two empirical user studies we demonstrate that our representation significantly outperforms both previous work and state-of-the art embedding baselines on these tasks. We achieve Mean Average Precision (MAP) of 87\% in the alternative product uses search, and 62\% of our inspirations for design space exploration are found to be \emph{useful and novel} -- a relative boost of 50-60\% over the best-performing baselines. 
\end{compactitem}

%% file: 02_related.tex
The contributions in this paper relate broadly to previous work on systems that use structured representations for supporting ideation, and studies that seek to understand the cognitive process of creativity and its implications. We provide a brief discussion of these two themes.

\xhdr{Cognitive Theories of Creative Thinking with Prior Ideas} 
A core aspect of creative thinking that distinguishes it from regular problem solving is the need for divergent thinking \cite{debonoLateralThinkingTextbook2010,sawyerExplainingCreativityScience2012,runcoDivergentThinking2019} - to construct and explore diverse ideas that are quite different from the obvious path of ideation. This process of divergence is shaped by prior knowledge \cite{wardWhatOldNew1995} --- such as past ideas, external stimulation, and examples --- in ways that sometimes hinders creativity through mechanisms like fixation \cite{janssonDesignFixation1991}, and sometimes leads to creative breakthroughs \cite{sioFixationInspirationMetaanalytic2015}.

Our design goals for this research are guided by past research on cognitive mechanisms that enable helpful interactions with past ideas. For example, research on insight problem-solving has uncovered the role of re-representation of past ideas by decomposing them into conceptual chunks and then recombining and/or repurposing them into new solutions  \cite{knoblichConstraintRelaxationChunk1999, mccaffreyInnovationReliesObscure2012}. Similar patterns in terms of core helpful interactions with prior ideas have been observed in in situ studies of expert designers' dissection of past ideas into component aspects and features for repurposing and recombination into new ideas \cite{eckertAdaptationSourcesInspiration2003, herringGettingInspiredUnderstanding2009, hargadonTechnologyBrokeringInnovation1997}. Another core process is analogical abstraction, where innovators think about and retrieve past ideas not in terms of their surface features, but in terms of deeper structural features or schemata, such as their underlying \textit{purposes} (goals) and \textit{mechanisms} \cite{gick1980analogical, gick_schema_1983}. These abstracted ``schemas" can then facilitate analogical transfer of ideas across domains that can lead to groundbreaking discoveries
\cite{hesse1966models,pask2003mathematics, markman2010structural, dahl2002influence,kittur2019scaling}. For example, the ancient Greeks studied the properties of sound waves by analogy to ripples in water; Nobel laureate Salvador Luria used abstract structural similarity between a slot machine and bacteria mutations to understand bacterial replication \cite{murray2016salvador}; and in computer science and optimization, analogies to processes in nature inspired algorithms such as simulated annealing \cite{kirkpatrick1983optimization}, genetic algorithms \cite{gen2007genetic}, and momentum-based gradient descent \cite{qian1999momentum}.  This process of abstraction over past ideas is also an important contributor to the ability to reformulate problems \cite{de1970lateral,dorstCoreDesignThinking2011,dubberlyModelingAnalysissynthesisBridge2008,kerne2014using,frich2019mapping}. For example, innovators tasked with a problem (find more room to store e-waste) might consider a related, more general goal (reducing environmental pollution) to inspire new solutions, or consider ``sibling'' formulations (create alternative materials that are biodegradable). This ability to reframe a problem using other problems that bear some abstract relation to it is known to be a powerful way to combat fixation and boost creativity \cite{gentner1997analogy,gentner1997structure,gentner_relational_2005,chan2011benefits,frich2019mapping}. This abstraction and reframing process has also been observed in studies of example curation \cite{kerneStrategiesFreeFormWeb2017, herringGettingInspiredUnderstanding2009}.

A core unifying thread across these mechanisms is the need for particular structured, yet flexible, representations of ideas in terms of their component aspects, such as analogical schemas, or conceptual chunks. In this paper, we focus on developing representations with these properties, starting with the decomposition of ideas into their component purposes and mechanisms. As we discuss in the next section, developing computational representations with these properties that can operate over large scale repositories of ideas remains a formidable open challenge.

\xhdr{Utilizing Structured Representations for Ideation Systems} 
A main focus of creativity techniques and prototypes has been building computational systems for exploring the space of possible solutions to problems and alternative problem perspectives \cite{frich2019mapping}. To do so, such systems often leverage structured knowledge representations for mapping the design space and linking across different problems. For example, the WordTree method \cite{linseyWordTreesMethodDesignbyanalogy2008} -- a prominent design method in creative engineering design -- directs designers to break their problem into subfunctions, and then use the WordNet database \cite{miller1995wordnet} to explore abstractions and related functional aspects. Likewise, a recent study \cite{gilon2018focab} asked designers to select product aspects to abstract using WordNet and the Cyc ontology \cite{lenat1995cyc}, which aimed to serve as a general-purpose repository of commonsense knowledge in structured form. These and other general-purpose knowledge bases (e.g., NELL \cite{mitchell2018never} and DBpedia \cite{farber2018linked}) largely encode categorical knowledge (e.g., is-a, has-a) and rarely functional knowledge (e.g., used-for), and often suffer from poor coverage of concepts in real-world products \cite{gilon2018focab}. Knowledge bases and ontologies that \emph{do} focus on functions, behaviors, and structures \cite{hirtz2002functional,altshuller200240,vattam_dane:_2011} have primarily been hand-crafted and are therefore even more limited in coverage. 
Work attempting to scale up has shown promise but is limited in expressivity or interpretability, such as modeling patents in terms of verbs and nouns \cite{fu_discovering_2013}, using principal component analysis \cite{duflouSystematicInnovationPatent2011a}, or learning coarse aggregate vectors that capture only one overall product purpose and mechanism \cite{hope2017accelerating} that cannot disentangle the different aspects of a product, unlike our work presented in this paper. While full abstraction of ideas currently remains a holy grail, here we investigate whether learning nuanced functional aspects might enable a limited form of abstraction useful for augmenting creativity and providing a first step towards true automated abstraction.

%% file: 02_general_formulation.tex
Our goal in this section is to construct a representation that can support the creative innovation tasks and interactions discussed in the Introduction. 

We propose to use \emph{span representations} \cite{kuribayashi-etal-2019-empirical}. Given a product text description, we extract tagged spans of text corresponding to purposes and mechanisms (see Figure \ref{fig:annot}), and represent the product as a set of span embeddings. 

More technically, we use a standard \emph{sequence tagging} formulation, with $\mathcal{X}_N = \{ \mathbf{x}_1, \mathbf{x}_2,\ldots, \mathbf{x}_N \}$ a training set of $N$ texts, each a sequence of tokens $\mathbf{x}_i = ({x}^1_i,{x}^2_i,\ldots,{x}^T_i)$, and $\mathcal{Y}_N$ a corresponding set of label sequences, $\mathcal{Y}_N = \{ \mathbf{y}_1, \mathbf{y}_2,\ldots, \mathbf{y}_N \}$, $ \mathbf{y}_i = \{ y^1_i, y^2_i,\ldots, y^T_i \}$, where each $y^j$ indicates token $j$'s label (purpose/mechanism/other). 
%\tnote{do we want to add here a more formal representation of a collection of span embeddings -- moving it up from section 3?}
%
In later sections, we represent each product $i$ as a set of purpose span embedding vectors and a set of mechanism span embedding vectors. %  $ \mathcal{M}_i \coloneqq \{\mathbf{p}^1_i, \mathbf{p}^2_i, \ldots, \mathbf{p}^{M_i}_i  \}$.

\subsection{Data}

\begin{figure}
    \includegraphics[width=\columnwidth]{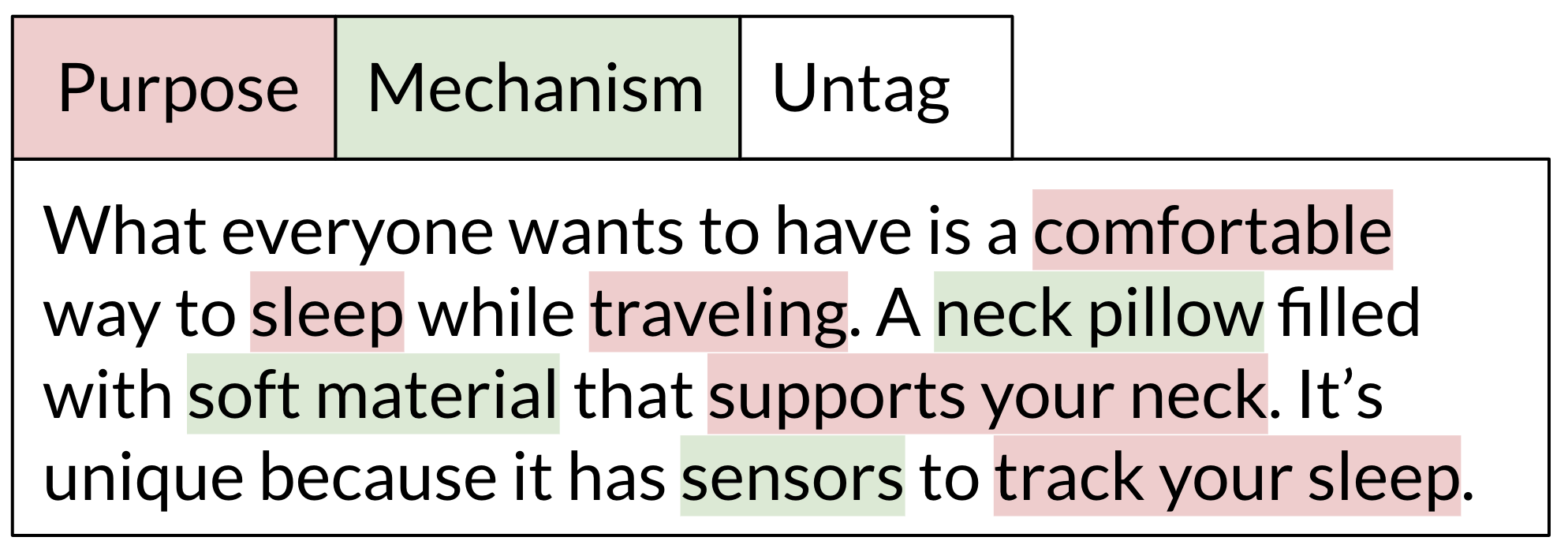}
    \caption{Crowdsourcing interface for fine-grained purposes and mechanisms.}
    \label{fig:annot} 
\end{figure}

We use real-world product idea descriptions taken from crowdsourced innovation website Quirky.com and used in \cite{hope2017accelerating}, including 8500 user-generated texts describing inventions across diverse domains (e.g., kitchen products, health and fitness, clean energy). Texts typically include multiple purposes and mechanisms. Texts in Quirky use very \textit{nonstandard language}, including grammatical and spelling errors (e.g., ``Folds Up Perfect For Carrying. you can walk-on, put your mouth on and or hands on. numbers in any configuration 4 learning to De / Composing Numbers.'').

\remove{
{\small 
\it 
 123 4FUN .
   An Educational Math Mat 4 Kids; 
   A mix between "Twister", "Simon Says", and "Whac-A-Mole". 
   A mat with Number Blocks. 
   A mat with numbers in any configuration 4 learning Place Values.
   Give commands like "what 2 numbers added together will give you 10".
   Hypo-allergenic.
   Have a whac-a-mole type mallet to hit the numbers.
   A mat with any shape and color with a number.
   Folds Up Perfect For Carrying.

}}

\remove{
{\small 
\it 
 An absorptive article containing a surface material comprising a
 combined non-woven fabric comprising at least two layers of a long
 fiber non-woven fabric and a short fiber non-woven fabric joined 
 together and an absorbing body for retaining a body fluid is disclosed
 in which the short fiber non-woven fabric is composed of 
 hot-melt-adhesive composite short fibers having at least two kinds of
 thermoplastic resin components of a high melting point component and
 low melting point component, and the hot-melt-adhesive composite short
 fibers are hot-melt-adhered together, the crossing angle of the short
 fibers at least preferably at least 45%, preferably at least 50% of
 the total contact points in the short fiber non-woven fabric are 
 occupied by an angle of 60 degree to 90 degree in the analysis of
 the distribution of the crossing angle at the contact points of 
 the fibers.
}
}

\xhdr{Annotation}\remove{In a recent NER competition on noisy text \cite{aguilar2017multi}, the winning entry reached average $F1$ of only $40\%-42\%$ -- far from the $93\%$ achieved on newswire sentences  \cite{akbik2018contextual}.}
To create a dataset annotated with purposes and mechanisms, we collect crowdsourced annotations on Amazon Mechanical Turk (AMT). In the similar annotation task of \cite{hope2017accelerating} workers were reported to annotate long, often irrelevant spans. We thus guided workers to focus on shorter spans. To further improve quality and encourage more fine-grained annotations, we limited maximal span length that could be annotated, and disabled the annotation of stopwords. 
Fig.~\ref{fig:annot} shows our tagging interface; rectangles are taggable chunks.
For quality control, we required US-based workers with approval rate over 95\% and at least 1000 approved tasks, and filtered unreasonably fast users. In total, we had $400$ annotating workers. Workers were paid \$0.1 per task. This rate was computed aiming for an hourly rate of $\$7$, where completion time was estimated via a small-scale pilot study. However, in the full study we were surprised to find the median completion time was much higher, reaching $100$ seconds. We note that this figure could be skewed (e.g., due to workers queuing of tasks or the ability to take breaks).

While a manual inspection of the annotations revealed they are mostly satisfactory,
 we observe two main issues: First, there are often {\bf multiple correct annotations}. Second, workers provide {\bf partial tagging} -- in particular, if similar spans appear in different sentences, very few workers bother tagging more than one instance (despite instructions). 
These issues would have made computing evaluation metrics problematic. 
We thus decided to use the crowdsourced annotations as a \emph{bronze-standard} for training and development sets only. For a reliable evaluation, we collected \emph{gold-standard} test sets annotated by two CS graduate students. Annotators were instructed to mark \emph{all} the relevant chunks, resulting in high inter-annotator agreement of $0.71$.  
We collect $22316$ annotated training sentences and $512$ gold sentences, for a total of $238,399 tokens$ (tag proportions: $14.5\%$ mechanism, $15.9\%$ purpose, $69.6\%$  other).

\xhdr{A note on related annotated data} 
There has been recent work on the related topic of information extraction from \emph{scientific papers} by classifying sentences, citations, or phrases. 
Recent supervised approaches \cite{jin2018hierarchical, augenstein2017semeval, luan2018multitask} use annotations which are often provided by either paper authors themselves, NLP experts, domain experts, or involve elaborate (multi-round) annotation protocols.
Sequence tagging models are often trained and evaluated on (relatively) clean, succinct sentences \cite{zhang2018graph,Marcheggiani2017}. When trained on noisy texts, results typically suffer drastically \cite{aguilar2017multi}. Our corpus of product descriptions is significantly noisier than scientific papers, and our training annotations were collected in a scalable, low-cost manner by non-experts.
Using noisy crowdsourced annotation for training and development only is consistent with our quest for a lightweight annotation approach that would still enable training useful models.

%% file: 03_acc_eval.tex
After collecting annotations, we can now train models to extract the spans. 
We explore several models likely to have sufficient power to learn our proposed novel representation, with the goal of selecting the best performing one. 
In particular, we chose two approaches that are common for related sequence-tagging problems, such as named entity recognition (NER) and part-of-speech (POS) tagging: a common baseline and a recent state-of-the-art model. We also tried a model-enrichment approach with syntactic relational inputs. We briefly describe the models we used below, with full technical descriptions and implementation details, data and code appearing in the supplementary material (Appendix \ref{subsec:app_modeldetails}). We note that our goal in this section is to find a \emph{reasonable model} whose output could support creative downstream tasks; many other architectures are possible and could be considered in future work.

{
\setdefaultleftmargin{0.5em}{0.5em}{}{}{}{}
\begin{compactitem}
\item \xhdr{BiLSTM-CRF} A BiLSTM-CRF \cite{huang2015bidirectional} neural network, a common baseline approach for NER tasks, enriched with semantic and syntactic input embeddings known to often boost performance \cite{zhang2018graph}. We adopt the ``multi-channel'' strategy as in \cite{zhang2018graph}, concatenating input word embeddings (pretrained GloVe vectors \cite{pennington2014glove}) with part-of-speech (POS) and NER embeddings. A conditional random field (CRF) model over the BiLSTM outputs maximizes the tag sequence log likelihood under a pairwise transition model between adjacent tags \cite{akbik2018contextual}.

\item \xhdr{Pre-trained Language Model (Pooled Flair)} A pre-trained language model \cite{akbik2019pooled} based on contextualized string embeddings, recently shown to outperform other powerful models such as BERT \cite{devlin2019bert} in NER and POS tagging tasks and achieve state-of-art results.

\item\xhdr{GCN} 
We also explore a model-enrichment approach with syntactic relational inputs. We employ a graph convolutional network (GCN) \cite{Kipf2016} over dependency-parse edges \cite{zhang2018graph}. GCNs are known to be useful for propagating relational information and utilizing syntactic cues \cite{zhang2018graph,Marcheggiani2017}. The linguistic cues are of special relevance and interest to us, as they are known to exist for purpose/mechanism mentions in texts \cite{fu_discovering_2013}.
\end{compactitem}
}

\begin{figure}[t!]
%\centering -> This is irrelevant because of the '.5\textwidth' as Mico advised below.
\begin{minipage}[t]{.45\textwidth}\vspace{0pt}
  \includegraphics[width=.9\columnwidth]{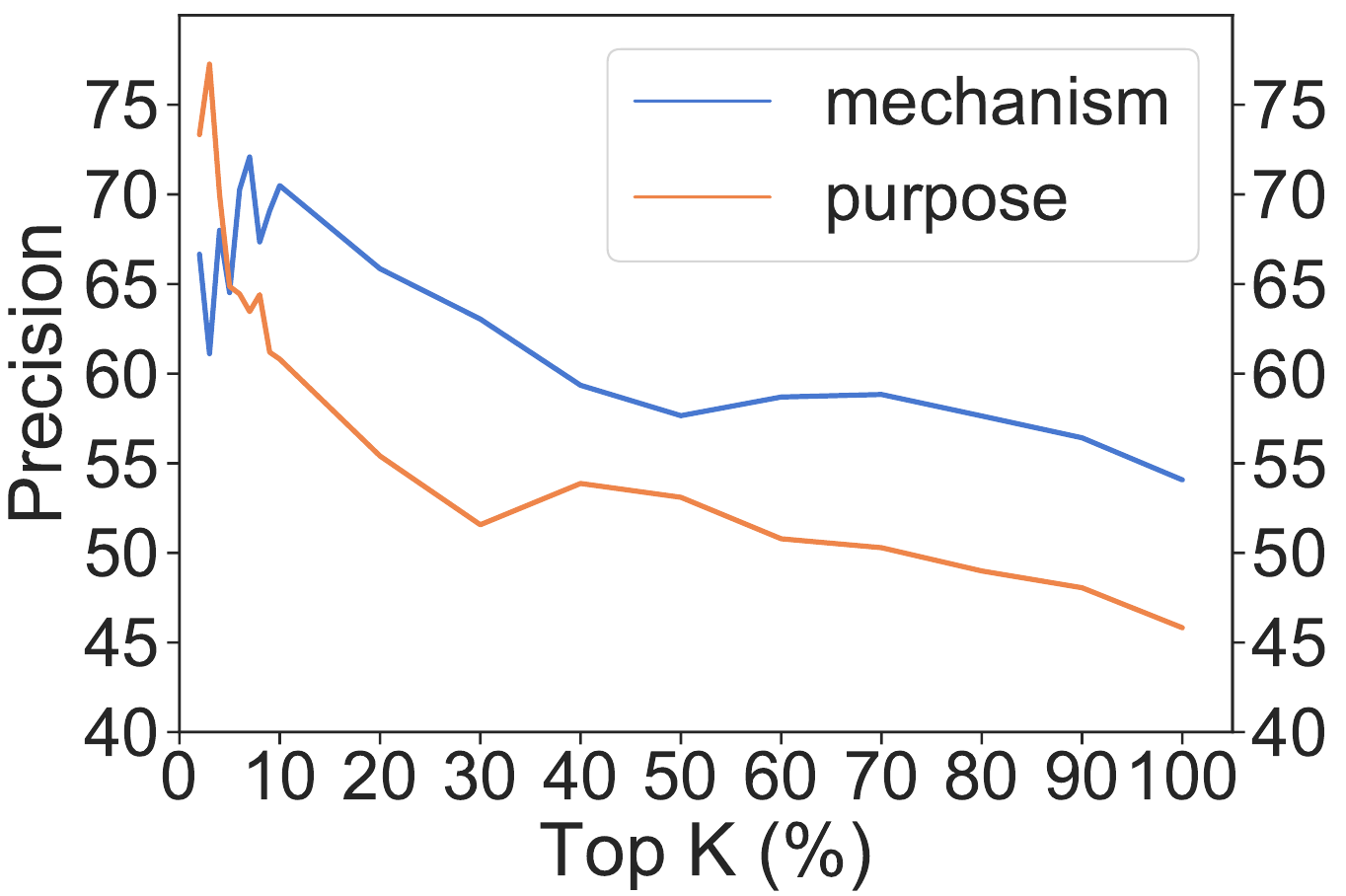}
\end{minipage}
\begin{minipage}[t]{.45\textwidth}\vspace{0.9cm}
\begin{tabular}[t]{llrrr}
\hline
\textbf{Configuration}  & \multicolumn{1}{c}{\textbf{P}} & \multicolumn{1}{c}{\textbf{R}} & \multicolumn{1}{c}{\textbf{$\Fone$}} \\ \hline
Enriched BiLSTM  &  45.24      & 39.01      & 41.90        \\
Pooled-Flair &  53.30      & 39.80       & 45.50       \\
GCN       &  47.85      & 47.93      & \textbf{47.89}     \\ \hline
%Syn+Sem GCN &  50.55      & 36.20       & 42.19       \\
GCN self-train  &  49.00      & 52.00      & \textbf{50.50} \\
\hline
\end{tabular}
\end{minipage}
\caption{Left: Precision@K results for the best performing model (GCN + self-training). Right: Raw extraction accuracy evaluation. All approaches use CRF loss. GCN with syntactic edges outperforms baselines. Self-training further improves results. Random-label achieves only $16.01$ $\Fone$. \label{fig:pk}} %\label{tab:results-quirky}
\end{figure}

\remove{
\begin{figure*}[t!]
 \begin{subfigure}[t]{0.45\textwidth}
    \includegraphics[width=\linewidth]{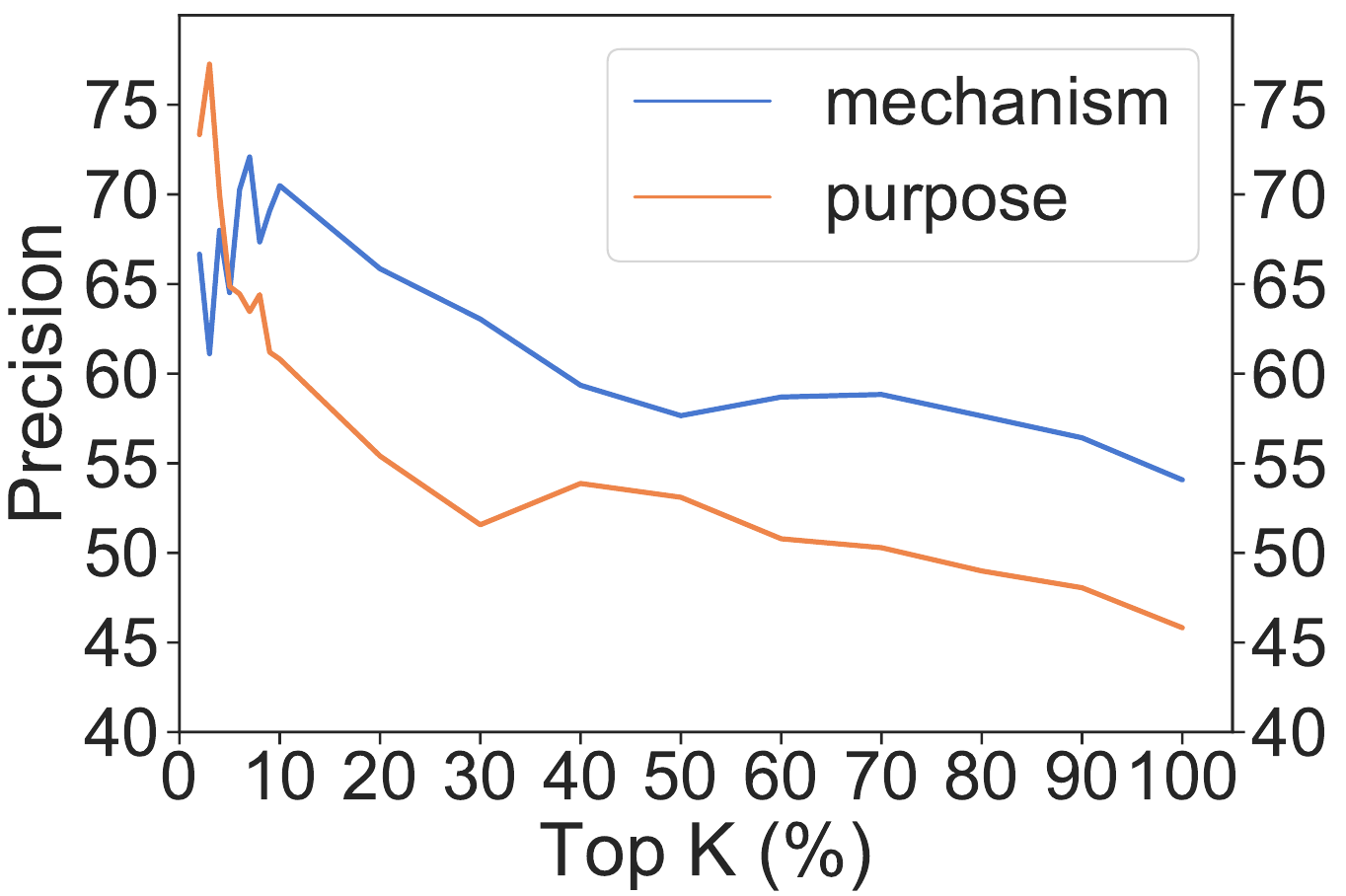}
    \caption{Precision@K results for the best performing model (GCN + self-training). %For low values of K, Pr@K is considerably higher than the overall precision. %For example, for K=5\%, precision is roughly 65\% for both purpose and mechanism predictions, 10-20 points higher than overall precision for mechanisms and purposes, respectively. 
    \label{fig:pk}}
\end{subfigure}
 \begin{subfigure}[t]{0.45\textwidth}
%\begin{adjustbox}{scale=.9,width=\columnwidth}
\begin{tabular}{llrrr}
\hline
\textbf{Configuration}  & \multicolumn{1}{c}{\textbf{P}} & \multicolumn{1}{c}{\textbf{R}} & \multicolumn{1}{c}{\textbf{$\Fone$}} \\ \hline
Enriched BiLSTM  &  45.24      & 39.01      & 41.90        \\
Pooled-Flair &  53.30      & 39.80       & 45.50       \\
GCN       &  47.85      & 47.93      & \textbf{47.89}     \\ \hline
%Syn+Sem GCN &  50.55      & 36.20       & 42.19       \\
GCN self-train  &  49.00      & 52.00      & \textbf{50.50} \\
\hline
\end{tabular}
%\end{adjustbox}
\caption{
}
\end{subfigure}
\end{figure*}
}

\subsection{Evaluation of Extraction Accuracy}
\label{sec:extacceval}

In this section we assess \emph{extraction accuracy} (whether we are able to extract purpose and mechanism spans of text). In the next sections, we evaluate the \emph{utility} of the extracted spans for enabling creative innovation tasks. 

To evaluate raw accuracy of the model's predictions, we use the standard IOB label markup to encode the \textit{purpose} and \textit{mechanism} spans (5 possible labels per token, \{Beginning, Inside\} x \{Purpose, Mechanism\} plus an "Outside" label). We conduct experiments using a train/development/test split of 18702/3614/512 \remove{and 8170/1253/350 sentences (for Quirky, patents respectively)}.

\begin{figure}
  %\remove{
%\begin{figure}[t]
    %\centering
    \includegraphics[width=0.95\columnwidth]{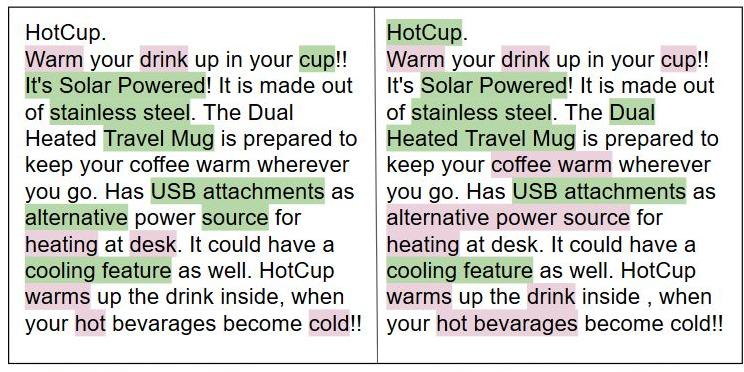}
    \caption{Comparing our GCN model predictions (right) to human annotations (left). Interestingly, our model managed to correct some annotator errors (``it's'', ``heated'', ``coffee warm'', ``beverages''). Purposes in pink, mechanisms in green.}
    \label{fig:models_predictions} 
%\end{figure}
%}
\end{figure}

Due to our challenging setting, we train models on bronze-standard annotations with noisy and partial tagging done by non-experts; for evaluation we use a curated gold-standard test set (Section \ref{formmodel-sec}). See Figure \ref{fig:pk} (right) for results: GCN reaches an $\Fone$ score of $\sim 48\%$, outperforming the BiLSTM-CRF model (enriched with multi-channel GloVe, POS, NER and dependency relation embeddings) by $6\%$. GCN also surpasses the strong Pooled-Flair pre-trained language model by nearly $2.5\%$. A random baseline guessing each token by label frequencies (Section \ref{formmodel-sec}) achieves $16.01$ $\Fone$. We interpret these results as possibly attesting to the utility of graph representations and 
features capturing syntactic and semantic information when labels are noisy. As a sanity check, we also computed precision@K (Figure \ref{fig:pk}, left). As expected, precision is higher with low values of $K$, and gradually degrades. Precision for mechanisms is higher than for purposes. Interestingly, a manual inspection revealed many cases where despite the noisy training setting, our models managed to correct mistaken or partial annotations (see Figure \ref{fig:models_predictions}).  

\xhdr{Self-Training} According to the results, we chose GCN as our best-performing model. We experimented adding self-training \citep{sachan2018self} to GCN. Self-training is a common approach in semi-supervised learning where we iteratively re-label ``O'' tags in training data with model predictions. A large portion of our training sentences are (erroneously) un-annotated by workers, perhaps due to annotation fatigue, introducing bias towards the ``O'' label. 

Self-training with GCN shows an improvement in $\Fone$ by an additional $2.6\%$, substantially increasing recall (more than $12\%$ over Flair), see Figure \ref{fig:pk}, right. Self-training stopped after $2$ iterations, following no gain in $\Fone$ on the development set. 

In the following two sections we demonstrate that our extraction model's accuracy, while far from perfect, is sufficient for achieving good performance on the \emph{downstream} tasks which are at the focus of this paper. One main reason for this gap is that our downstream tasks involve aggregation of multiple extracted spans: Product descriptions will typically mention salient mechanisms/purposes several times in the text, such that the effect of local false positives/negatives is mitigated if overall the key aspects are captured somewhere in the text. Further, as we discuss in §\ref{sec:fcg}, our approach also aggregates purposes and mechanisms across the \emph{entire corpus}, not just single texts, learning from patterns observed sufficiently many times across multiple texts and thus removing noise introduced by extraction errors. As future information extraction technologies advance, our task could benefit from improved extraction accuracy to further reduce the rate of false positives and negatives.

\remove{
\begin{table}
\centering
%\begin{adjustbox}{width=\columnwidth}
\begin{tabular}{lll}
Class Label / Dataset & Quirky \%  & Patents \% \\
\hline
Mechanism (M)       & 14.45 & \hphantom{0}8.06  \\
Purpose (P)        & 15.94 & 10.13 \\
Other (O)           & 69.61 & 81.81 \\
\hline
Total Count           & 238,399 & 288,929
\end{tabular}
%\end{adjustbox}
\caption{\label{tab:dataset-stats}Label type distribution for Quirky and Patents datasets. Note the high proportion of ``O'' labels due to partially annotated sentences.}

\end{table}
}

%% file: 04_Search.tex
\section{Fine-Grained Functional Search for Alternative Uses}
\label{sec:search}

In the previous section we suggested a model for extracting purpose and mechanism spans and assessed extraction accuracy.  
Our focus in this paper is to study the \emph{utility} of the extracted purposes and mechanisms, in terms of the user interactions they enable. In the following sections we explore two tasks demonstrating the value of our novel representation for supporting creative innovation.  We start with a task involving search for \emph{alternative uses}. 

\xhdr{Motivation} Our task is inspired by one of the most well-known divergent thinking tests  \cite{guilford1959three} for measuring creative ability -- the alternative uses test \cite{guilford1967nature}, where 
participants are asked to think of as many uses as possible for some object. Aside from serving as a measure of creativity, the ability to find alternative uses for technologies has important applications in engineering, science and industry. Technologies developed at NASA, the US space agency, have led to over 2,000 spinoffs, finding new uses in computer technology, agriculture, health, transportation, and even consumer products\footnote{\url{https://spinoff.nasa.gov/}}. Procter \& Gamble, the multinational consumer goods company, has invested in systematic search for ideas to re-purpose and adapt from other industries, such as using a compound that speeds up wound healing to treat wrinkles - an idea that led to a new line of anti-wrinkle products \cite{pg}. And very recently, the COVID-19 pandemic provided a stark example of human innovation, with many companies seeking to pivot and re-purpose existing products to fit the new climate \cite{covid}.

One teaching story is that of John Osher, creator of the ``Spin Pop'' --- a lollipop with a mechanism for twirling in your mouth. After selling his invention, Osher's team systematically searched for new ideas --- ``rather than having an idea come to us''\footnote{\url{https://www.allbusiness.com/the-man-the-legend-john-osher-inventor-of-the-spin-brush-part-i-2-7665547-1.html}}. The group eventually landed on the ''Spin Brush'' -- a cheap electric toothbrush adapted from the same twirling mechanisms. This case of repurposing an existing technology involved a systematic search process rather than pure serendipity. % which required a rich, fine-grained understanding of products  %However, Osher and his team still had to rely on human processing power -- inherently limited in its ability to scour millions of potential descriptions of problems available online and find relevant problems.
Introducing automation could help accelerate the search process by scouring many relevant problems available online, but the task is challenging for existing search systems, requiring a fine-grained, multi-aspect understanding of products.

\xhdr{Illustrative Example} Consider a company that manufactures light bulbs. The company is familiar with straightforward usages of their product (lamps, flashlights), and wants to identify non-standard uses and expand to new markets. Finding uses for a lightbulb that are not about the standard purpose of illuminating a space would be difficult to do with a standard search query over an idea repository, as the term ``lights'' or ``lighting'' will bring back lots of results close to ``lamps,'' ``flashlight,'' and the like. 
In contrast, with our representation each idea is associated with mechanism and purpose aspects, and one could form a query such as \emph{mechanism=``light bulb''}, \emph{purpose= NOT ``light''}. Using our system, 
the searcher adds ``light'' as a mechanism and also adds ``light'' as a negative purpose (i.e., results should not include ``light'' as a purpose). Our prototype returns examples such as billiard laser instructor devices (Table~\ref{tab:searchres_ex}), warning signs on food packages to get attention of kids with allergies, and lights attached to furniture to protect your pinky toes at night (Figure~\ref{fig:light}).

\subsection{Study Design}
We have built a  search engine prototype supporting our representation. Figure \ref{fig:light} shows the top two results for the light bulb scenario: warning lights on food for kids with allergies, and lights attached to furniture to protect your pinky toe at night. These are non-standard recombinations \cite{fleming2001recombinant} (light + allergies, light + furniture guard) that could lead the company to new markets. 

We conduct an experiment simulating scenarios where users wish to find novel/uncommon uses of mechanisms. Table \ref{tab:searchres_ex} shows the scenarios and examples. To choose these scenarios for the experiment, we find popular/common mechanisms in the dataset and their most typical uses. For example, one frequent mechanism is RFID, which is typically used for purposes such as ``locating'' and ``tracking''. We then create queries searching for \emph{different} uses -- purposes that do not include concepts related to the typical uses of a given mechanism. To automate scenario selection, we cluster mechanisms (see Section \ref{hier-sec}), select frequent mechanisms from the top $5$ largest mechanism clusters, and identify purposes strongly co-occurring with them (e.g., ``RFID'' co-occurs with ``locating'', ``tracking'') to avoid.

\begin{figure}[t!]  
      \begin{minipage}{0.95\linewidth}
        \includegraphics[width=0.9\linewidth]{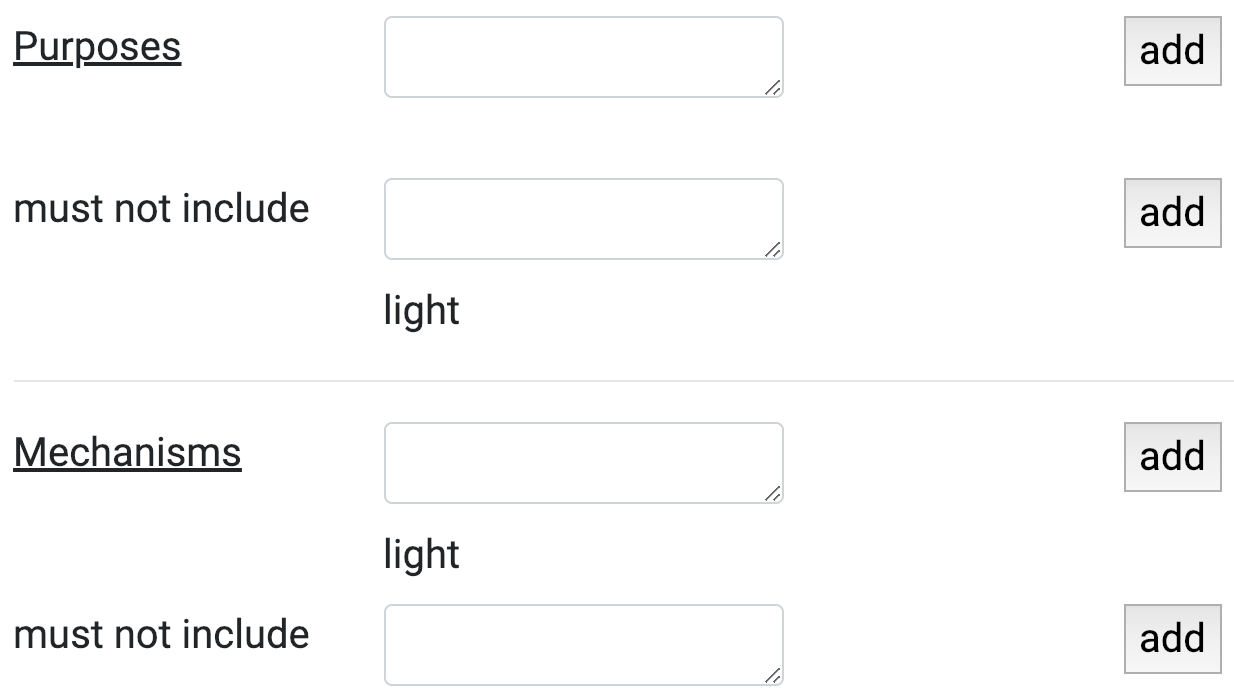}
        \end{minipage}
        %\ 
       
        \begin{minipage}{0.95\linewidth}
        \includegraphics[width=.9\linewidth]{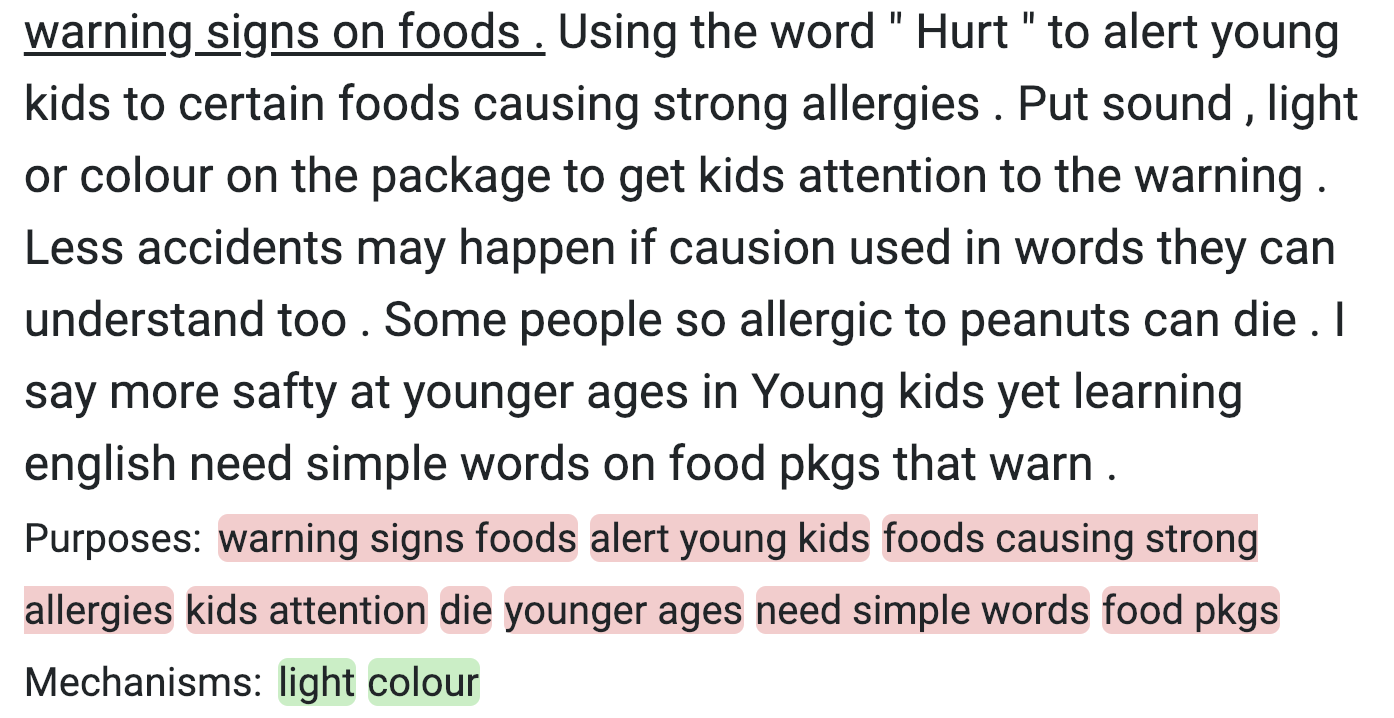}
    
     \  \hspace{0.2cm}
    
        \includegraphics[width=.9\linewidth]{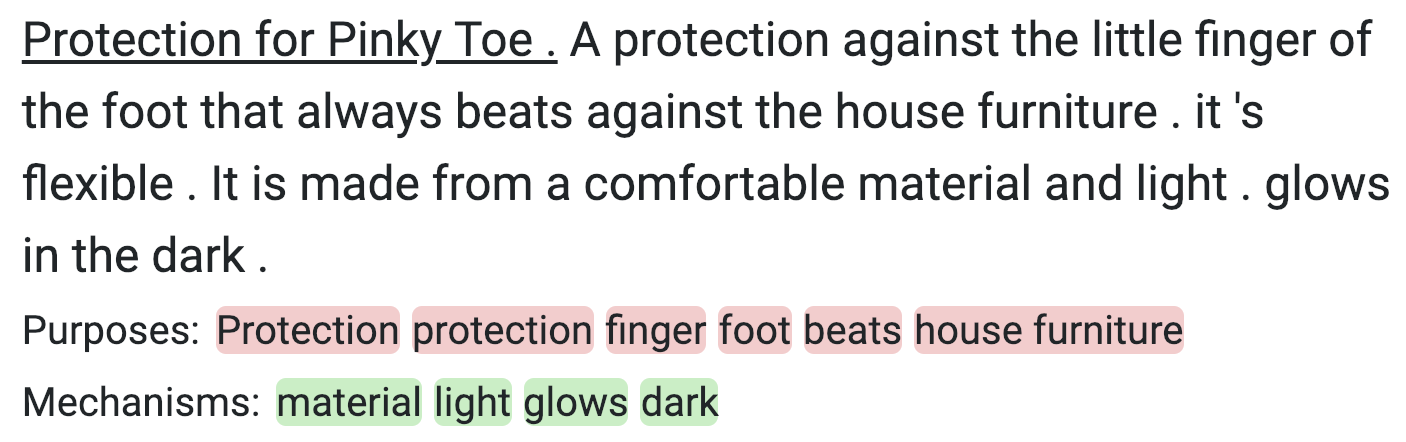}
        \end{minipage}
    \caption{Applications for light where light is not in the purpose. Left: Interface. Right: Two of the results and their automatic annotations (purposes in pink, mechanisms in green).\label{fig:light}}
\end{figure}

\remove{
\begin{figure}[t!]
    \centering
        \centering
        \fbox{\parbox{0.4\linewidth}{
        \includegraphics[width=0.9\linewidth]{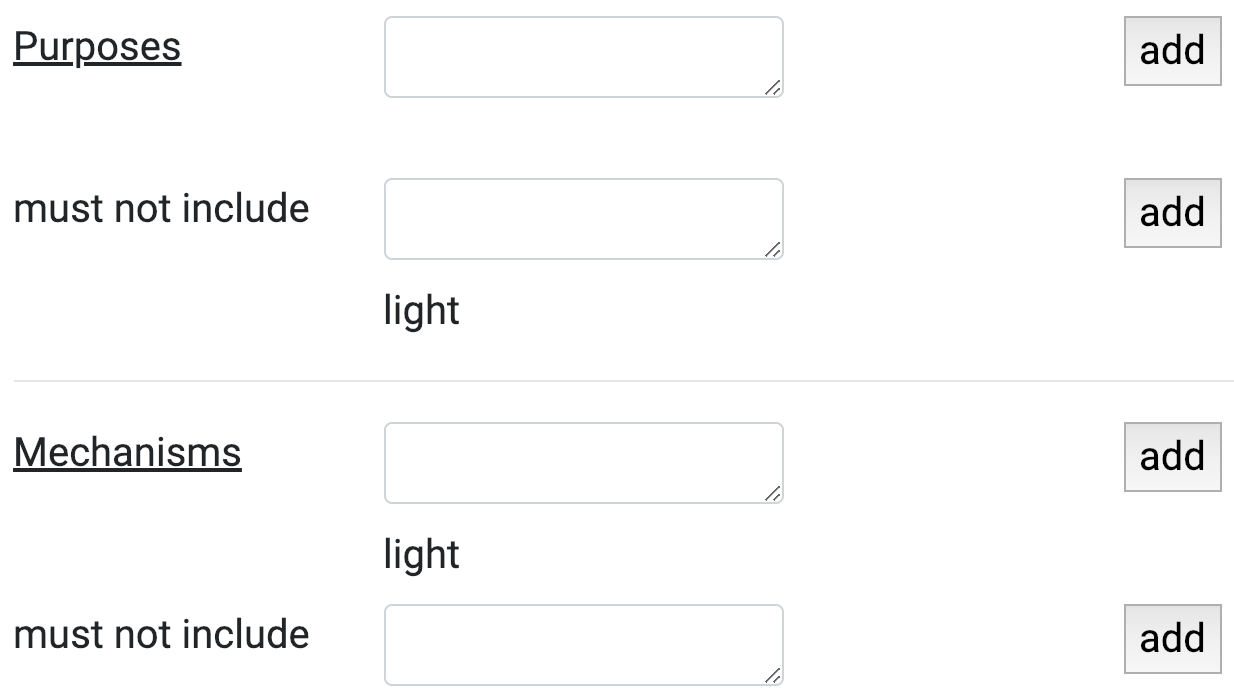}
        %
        %\ 
        %
        \includegraphics[width=\linewidth]{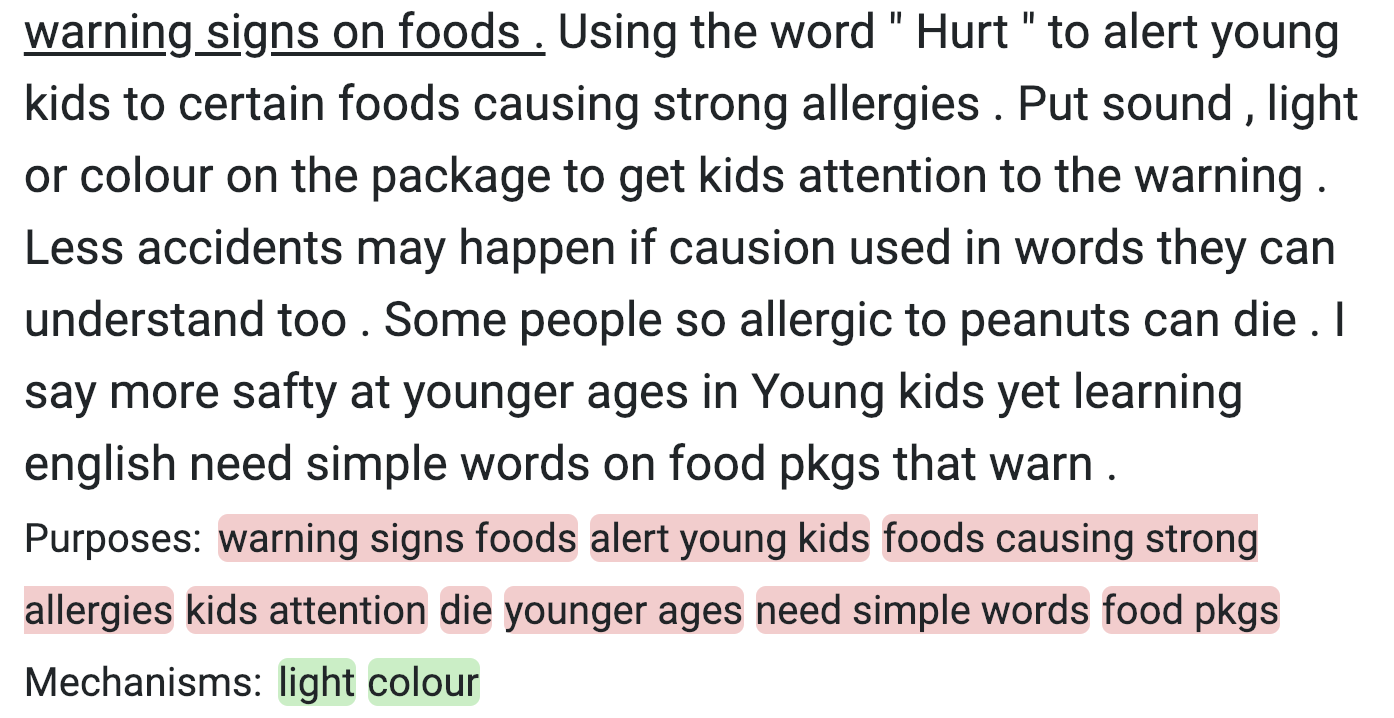}
    
     \  
    
        \includegraphics[width=\linewidth]{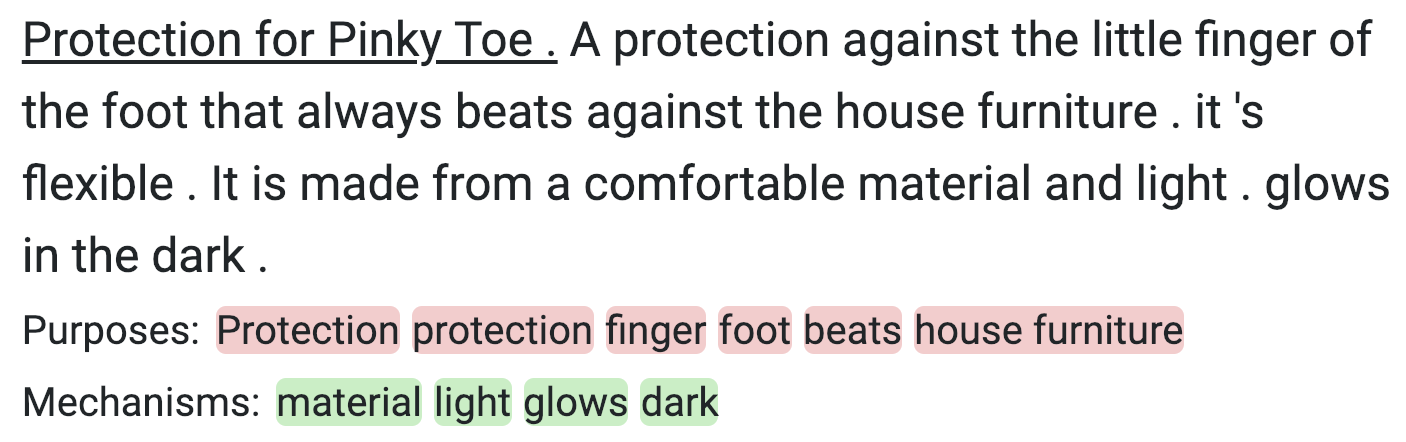}}}
    \caption{Applications for light where light is not in the purpose. Two of the results and their automatic annotations (purposes in pink, mechanisms in green).\label{fig:light}}
\end{figure}
}

\remove{
\begin{figure}
    \centering
    \begin{subfigure}[b]{\linewidth} 
        \centering
        \frame{\includegraphics[width=0.9\linewidth]{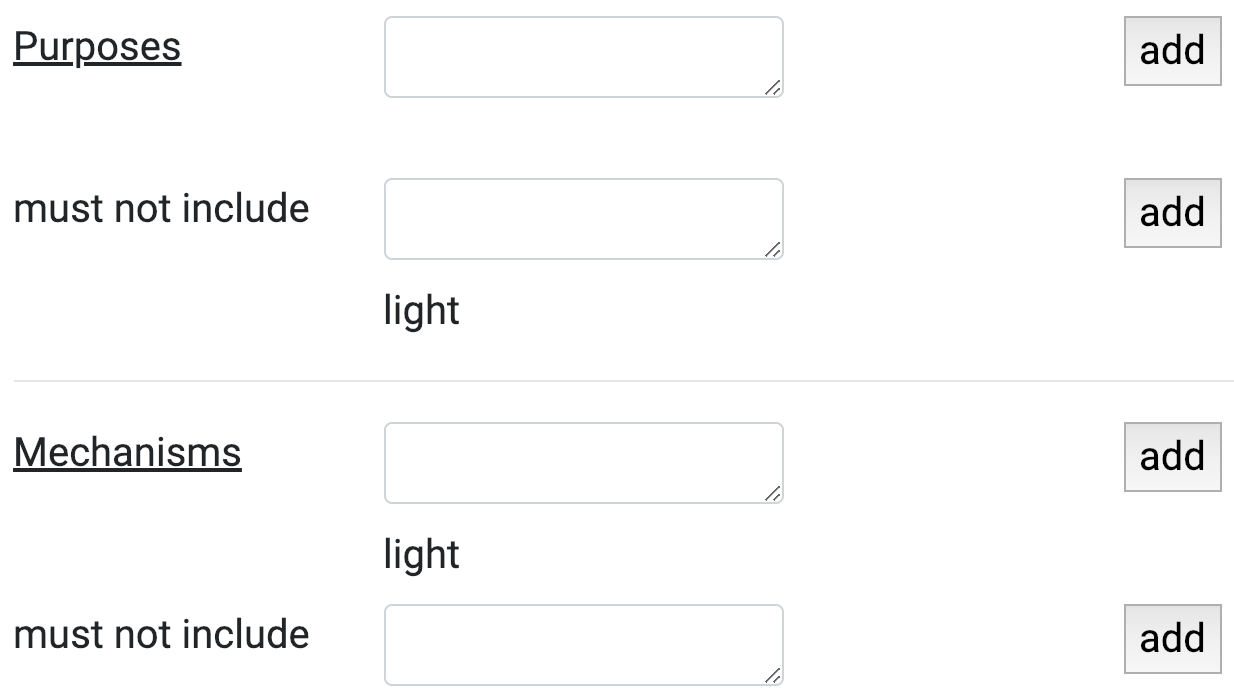}}
        \vspace{2mm}
    \end{subfigure}\hfill
    \begin{subfigure}[b]{\linewidth} 
        \centering
        \includegraphics[width=\linewidth]{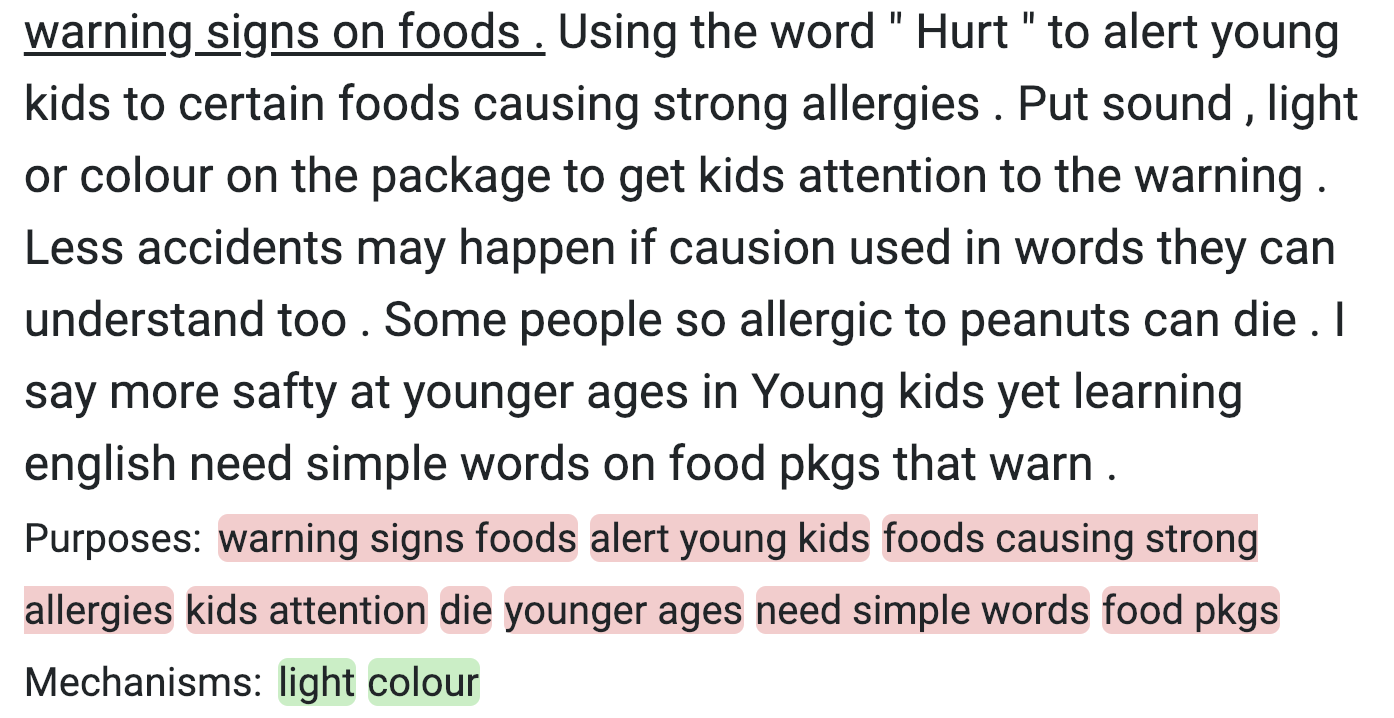}
    \end{subfigure}\hfill
    \begin{subfigure}[b]{\linewidth}
        \centering
        \includegraphics[width=\linewidth]{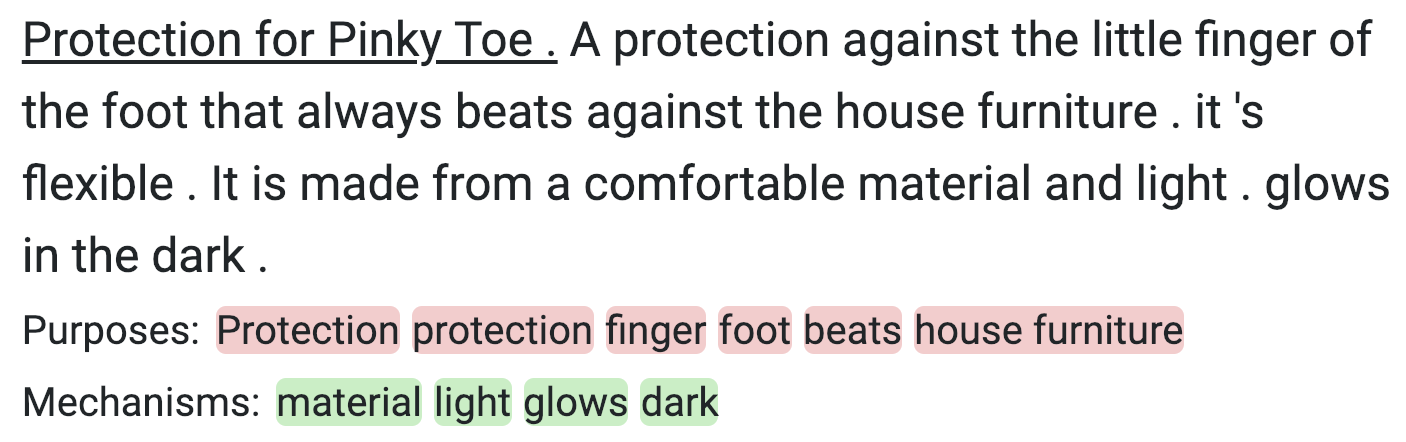}
    \end{subfigure}
    \caption{Applications for portable, detachable lights (detected purposes and mechanisms are presented in red and green spans, respectively). Top: search interface. Query: ``light as a mechanism, light cannot be in the purpose''. Bottom: two of the results and their (automatic) annotations.\label{fig:light}}
\end{figure}
}

\remove{
\begin{figure}
    \centering
    \begin{subfigure}[b]{\linewidth} 
        \centering
        \includegraphics[width=\linewidth]{rfid-checkout.png}
    \end{subfigure}\hfill
    \begin{subfigure}[b]{\linewidth}
        \centering
        \includegraphics[width=\linewidth]{rfid-luglock.png}
    \end{subfigure}
    \caption{Non-trivial applications of RFIDs: checkout at supermarkets or as luggage locks.\label{fig:rfid}}
\end{figure}
}

%% file: 05_search_eval.tex
%We base our formulation on \citet{hope2017accelerating}. In their framework, each product $\mathbf{x}_i$ had one vector representing its overall purpose, $\mathbf{p}_i$,  and one vector representing its overall mechanism $\mathbf{m}_i$. Then, $d_p(\cdot,\cdot)$ and $d_m(\cdot,\cdot)$
%were distance metrics for purpose vectors and for mechanism vectors, respectively.

%Denoting by  and  the purpose and mechanism vectors for product $i$, the authors proposed applying rich queries to a database of product texts $\mathcal{D}$. Let $d_p(\cdot,\cdot), d_m(\cdot,\cdot)$ be distance metrics over the vectors. Given these distances one could frame core analogy tasks as optimization problems. For example, query to retrieve products solving similar problems to a given product but with different mechanisms was formulated as:
\remove{
\begin{compactitem}
\item \textit{Same purpose, different mechanism} :
\begin{equation}
\label{eq:same-purp}
\begin{aligned}
\underset{\mathbf{\tilde{i}}\in \mathcal{P} }{\text{argmin }} &   
d_p(\mathbf{p}_i,\mathbf{p}_{\tilde{i}}) \\
s.t.
&  d_m(\mathbf{m}_i,\mathbf{m}_{\tilde{i}}) \geq \text{threshold},
\end{aligned}
\end{equation}
\end{compactitem}
}

%However, real products are composed of \emph{multiple} purposes and mechanisms, often in varying levels of abstraction.

%In this section, we describe our approach to learning to extract purposes and mechanisms from products. We start by formally defining our general setting.

%We dive deeper into the problem formulation and modelling in Section \ref{intro-sec} \dnote{not in intro?}, and discuss how we extract the sets $\mathcal{P}_i$ and $\mathcal{M}_i$ for a product $i$. For now, assume our product representation consists of these two sets. 
%To allow the rich analogical mapping and search discussed above,  
\xhdr{Our Approach} We represent each product $i$ as a set of purpose vectors $\mathcal{P}_i \coloneqq \{\mathbf{p}^1_i, \mathbf{p}^2_i, \ldots, \mathbf{p}^{P_i}_i  \}$, and a set of mechanism vectors  $ \mathcal{M}_i \coloneqq \{\mathbf{p}^1_i, \mathbf{p}^2_i, \ldots, \mathbf{p}^{M_i}_i  \}$ extracted with our GCN model.
%The elements are vectors corresponding to purpose and mechanism tokens. 
%, and $P,M$ are the sizes of each purpose/mechanism set, respectively.
%Each vector $\mathbf{p}^k_i$ or $\mathbf{m}^m_i$ \dnote{did you really mean $m^m$?} 
Similarly, we define a set of query vectors $\mathbf{q}_p \coloneqq {\mathbf{q}_1, \mathbf{q}_2, \ldots \mathbf{q}_{Q_p}}$ and  $\mathbf{q}_m \coloneqq {\mathbf{q}_1, \mathbf{q}_2, \ldots \mathbf{q}_{Q_m}}$. Each query chunk can be negated, meaning it should not appear. Finally, we define distance metrics $d_p(\cdot,\cdot)$, $d_m(\cdot,\cdot)$ between \emph{sets} of purposes and mechanisms.
For example, to locate a dog using RFID but \textit{not} GPS: 
%$\mathcal{P}_i$ and $\mathcal{M}_i$ as their respective arguments for product $i$. Finally, we define the sets of query vectors $\mathbf{q}_p \coloneqq {\mathbf{q}_1, \mathbf{q}_2, \ldots \mathbf{q}_{Q_p}}$ and  $\mathbf{q}_m \coloneqq {\mathbf{q}_1, \mathbf{q}_2, \ldots \mathbf{q}_{Q_m}}$ for purpose and mechanism queries. In addition, each query chunk can be negated (meaning it should not be included). Using this formulation, we can support expressive queries.%
\begin{equation}\label{eq:same-purp}
\begin{aligned}
%\nonumber
{\text{argmin }_i} \ \ &   
d_p(\{\mathbf{q}_\text{``locate dog''}\}, \mathcal{P}_{\tilde{i}}) \\
s.t.\ \ 
&  d_m(\{\mathbf{q}_{\text{``GPS''}}\},\mathcal{M}_{\tilde{i}}) \geq \text{threshold} \\
&  d_m(\{\mathbf{q}_{\text{``RFID''}}\},\mathcal{M}_{\tilde{i}}) \leq \text{threshold}
\end{aligned}
\end{equation}

%We can now formulate a rich, fine-grained query capturing multiple purpose-mechanism conditions. For example, consider a product engineer wishing to search for ideas that solve the problems of tracking lost children and locating misplaced wallets, using an RFID mechanism but NOT a GPS mechanism. Our framework supports such rich queries, as follows: Let
%$\mathbf{q}_p$ may correspond to vector embeddings of the purpose query terms [\textit{locating misplaced wallets, tracking lost children}] and let the mechanism queries $\mathbf{q}_m$  correspond to [\textit{GPS, RFID}]. \dnote{I changed some, not sure if that's what you meant. also not clear if semantics of a set are AND or OR. plus some are negative. would you rather define a search expression over the qs?} We can formulate the problem as

\remove{
\begin{compactitem}
\item \textit{Multi purpose match, including one mechanism but not another} :
\begin{equation}
\label{eq:same-purp}
\begin{aligned}
\underset{\mathbf{\tilde{i}}\in \mathcal{D} }{\text{argmin }} &   
d_p(\mathbf{q}_p, \mathcal{P}_{\tilde{i}}) \\
s.t.
&  d_m(\{\mathbf{q}_{\text{GPS}}\},\mathcal{M}_{\tilde{i}}) \geq \text{threshold} \\
&  d_m(\{\mathbf{q}_{\text{RFID}}\},\mathcal{M}_{\tilde{i}}) \leq \text{threshold},
\end{aligned}
\end{equation}
\end{compactitem}
}

%In \citet{hope2017accelerating}, the pre-trained GloVe \cite{pennington2014glove} word vectors for tokens ${x}^j_i$ corresponding to purpose tokens(mechanism tokens, respectively) were averaged together with TF-IDF weighting, resulting in two soft aggregate representations.

%We now turn to evaluating the ability of our framework to express rich analogical search queries in a user study, as in the example scenario in the previous section.
%\xhdr{Query scenarios} 

%This results in the following scenarios:

\begin{table*}
{\small
\begin{tabular}{p{0.5\linewidth}|p{0.4\linewidth}}
    \hline
    \textbf{Query} & \textbf{Example results}\\
    \hline

    \parbox{\linewidth}{Mechanism: \textit{light}. Purpose: NOT \textit{light}}
    %{Use \textit{light} for a purpose that is NOT mainly about \textit{lighting}} 
    &\parbox{\linewidth}{Billiard laser instructor (projector)}\\
    \hline    
    \parbox{\linewidth}{Mechanism: \textit{solar energy}. Purpose: NOT \textit{generating power}} &\parbox{\linewidth}{Light bulbs with built-in solar chips.}\\
    \hline
    \parbox{\linewidth}{Mechanism: \textit{water}. Purpose: NOT \textit{cleaning}, NOT \textit{drinking} } &\parbox{\linewidth}{A lighter that burns hydrogen generated from water and sunlight.}\\
    \hline
    \parbox{\linewidth}{Mechanism: \textit{RFID}. Purpose: NOT \textit{locating}, NOT \emph{tracking}} &\parbox{\linewidth}{A digital lock for your luggage with RFID access.}\\
    \hline
    \parbox{\linewidth}{Mechanism: \textit{light}. Purpose: \textit{cleaning}} 
    &\parbox{\linewidth}{A UV box to clean and sanitize barbells at the gym.}\\
    \hline
\end{tabular}
}
\caption{Scenarios and example results retrieved by our FineGrained-AVG method. Queries reflect non-trivial uses of mechanisms (e.g., using water not for drinking/cleaning, retrieving a lighter running on hydrogen from water and sunlight).}
\label{tab:searchres_ex}
\end{table*}

\remove{
\begin{compactitem}
\item Mechanism: \textit{light}. Purpose: NOT \textit{light}
\item Mechanism: \textit{solar energy}. Purpose: NOT \textit{generating power}
\item Mechanism: \textit{water}. Purpose: NOT \textit{cleaning}, NOT \textit{drinking} 
\item Mechanism: \textit{RFID}. Purpose: NOT \textit{locating}, NOT \emph{tracking}
\item Mechanism: \textit{light}. Purpose: \textit{cleaning}
\end{compactitem}
}

\remove{
\begin{compactitem}
    \item You have a factory that makes components that use \textit{solar energy}. You want to come up with ideas for products that use solar energy, whose main goal is NOT \textit{generating power} (as in, you already thought of generators, chargers…).
    \item You have a factory that makes light-based products (lightbulbs, LED lights…). You want to come up with products that use \textit{light} for a purpose that is not mainly about \textit{light} (you already thought of lamps and flashlights)
    \item You want to come up with products that use \textit{light} for \textit{cleaning}.
    \item You want to use \textit{water} for a purpose that is not mainly \textit{cleaning} and \textit{drinking}.
    \item You want to use \textit{RFID} for products whose main goal is NOT locating and tracking things \textit{locating and tracking things}.
\end{compactitem}
}

\remove{
{ \setdefaultleftmargin{2em}{3em}{}{}{}{}
\begin{compactitem}
\item \xhdr{Scenario I}: Products that use \textit{solar energy}, whose main goal is NOT \textit{generating power}
\item \xhdr{Scenario II}: Use \textit{light} for the  purpose of \textit{cleaning}.
\item \xhdr{Scenario III}: Use \textit{water} for a purpose that is NOT mainly about \textit{cleaning} and \textit{drinking}.  
\item \xhdr{Scenario IV}: Use \textit{RFID} for products whose main goal is NOT \textit{locating and tracking things}.
\end{compactitem}}

We model a product $i$ as a set of purpose and mechanism vectors  $\mathcal{P}_i \coloneqq \{\mathbf{p}^1_i, \mathbf{p}^2_i, \ldots, \mathbf{p}^{P_i}_i  \}$, and  $ \mathcal{M}_i \coloneqq \{\mathbf{p}^1_i, \mathbf{p}^2_i, \ldots, \mathbf{p}^{M_i}_i  \}$.

For our chosen scenarios, we construct query sets for purpose and mechanism,  $\mathbf{q}_p \coloneqq {\mathbf{q}_1, \mathbf{q}_2, \ldots \mathbf{q}_{Q_p}}$ and  $\mathbf{q}_m \coloneqq {\mathbf{q}_1, \mathbf{q}_2, \ldots \mathbf{q}_{Q_m}}$. Certain elements in $\mathbf{q}_p$ or  $\mathbf{q}_m$ may correspond to \textit{negative queries}, meaning we wish to retrieve products far from those purposes or mechanisms (such as finding a product NOT using an RFID mechanism).
}

\noindent We explore two alternatives for computing distance metrics $d_m, d_p$:

%\dnote{can't we come up with better names? nobody will get PM unless they actually think about it}

{ \setdefaultleftmargin{1em}{2em}{}{}{}{}
\begin{compactitem}
\item \xhdr{FineGrained-AVG} $d_p(\mathbf{q}_p, \mathcal{P}_i)$
%= 1 - \frac{\Sigma^{Q_p}_{j=1}\mathbf{q}_j}{Q_p Z_q} \cdot \frac{\Sigma^{P_i}_{j=1}\mathbf{p}^j_i}{P_i Z_p}$, 
is 1 minus the dot product between average query and purpose vectors (normalized to unit norm). %respectively, after normalizing  (normalizing constants $Z$ are the $l_2$ norm of each average vector). 
We define $d_m$ similarly. 
%(\mathbf{q}_m, \mathcal{M}_i) = 1 - \frac{\Sigma^{Q_p}_{j=1}\mathbf{q}_j}{Q_p Z_q} \cdot \frac{\Sigma^{M_i}_{j=1}\mathbf{m}^j_i}{M_i Z_p}$. 

\item \xhdr{FineGrained-MAXMIN} We match each element in  $\mathbf{q}_p$ with its nearest neighbor in $\mathcal{P}_i$, and then find the minimum over the distances between matches. $d_p$ is defined as 1 minus the minimum. 
%\dnote{something is weird about your formula here, no?} %: $d_p(\mathbf{q}_p, \mathcal{P}_i) = 1- \min\limits_{j}\max\limits_{\mathbf{p} \in \mathcal{P}} \mathbf{p} \cdot \mathbf{q}_j $, where 
All vectors are normalized. We define $d_m$ similarly.  %$\mathbf{p}^j$ and $\mathbf{q}_j$ are assumed normalized (we drop indices for brevity of notation). $d_p(\mathbf{q}_m, \mathcal{M}_i)$ is defined in similar fashion. 
This captures cases where queries match only a small subset of product chunks, erring on the side of caution with a max-min approach. % to avoid spurious matches with only one $\mathbf{p}^j$ or $\mathbf{m}^j$ element in the text while the rest of the query terms are far from $\mathcal{P}$ and $\mathcal{M}$.

\end{compactitem}}

%The vectors $\mathbf{q}_p, \mathbf{q}_m$ are represented using pretrained GloVe vectors from \cite{pennington2014glove}. We also use this representation for $\mathbf{p}, \mathbf{m}$. While we could embed all query and purpose/mechanism terms into a shared space induced by our model in Section \ref{formmodel-sec}, we test how useful and generalizable the actual extracted textual tags are on their own without using representations fine-tuned to a specific data set.

\xhdr{Baselines} We test our model against:

{ \setdefaultleftmargin{1em}{1em}{}{}{}{}
\begin{compactitem}
\item \xhdr{AvgGloVe} A weighted average of GloVe vectors of the entire text (excluding stopwords), similar to standard NLP approaches for retrieval and textual similarity. We average query terms and normalize to unit norm. Distance is computed via the dot product.

\item \xhdr{Aggregate purpose/mechanism} Representing each document with the model in \cite{hope2017accelerating}. This model takes raw text as input, applies a BiLSTM neural network and produces two vectors corresponding to \emph{aggregate} purpose and mechanism of the document. We average and normalize query vectors, and use the dot product.

\end{compactitem}}

For all four methods, we handle negative (purpose) queries by filtering out all products whose similarity is lower than $\lambda$, where lambda is a threshold selected to be the $90^{\text{th}}$ percentile of similarities (1 minus the distances). This corresponds to the threshold seen in the example in Eq. \ref{eq:same-purp}.

\remove{\begin{compactitem}
    \item Products that use \textit{solar energy}, whose main goal is NOT \textit{generating power}. 
    \item Use \textit{water} for a purpose that is NOT mainly about \textit{cleaning}
    \item Use \textit{light} for the  purpose of \textit{cleaning}.  
    \item Use \textit{RFID} for products whose main goal is NOT \textit{locating and tracking things}.
\end{compactitem}
}

%    \item Mechanism: \textit{light}. Purpose: NOT \textit{light}
%\item Mechanism: \textit{solar energy}. Purpose: NOT \textit{generating power}
%\item Mechanism: \textit{water}. Purpose: NOT \textit{cleaning}, NOT \textit{drinking} 
%\item Mechanism: \textit{RFID}. Purpose: NOT \textit{locating}, NOT \emph{tracking}
%\item Mechanism: 

\subsection{Results}
We recruited five engineering graduate students (three female, two male) to judge the retrieved product ideas. 
Each participant provided binary relevance feedback \cite{schutze2008introduction} (yes/no) to the top 20 results from each of the four methods, shuffled randomly so that judges are blind to the condition.\footnote{Inter-rater agreement measured across all scenarios was at $50\%$ by both Fleiss kappa and Krippendorff's alpha tests.}

\begin{figure}
    \centering
    \includegraphics[width=0.9\columnwidth]{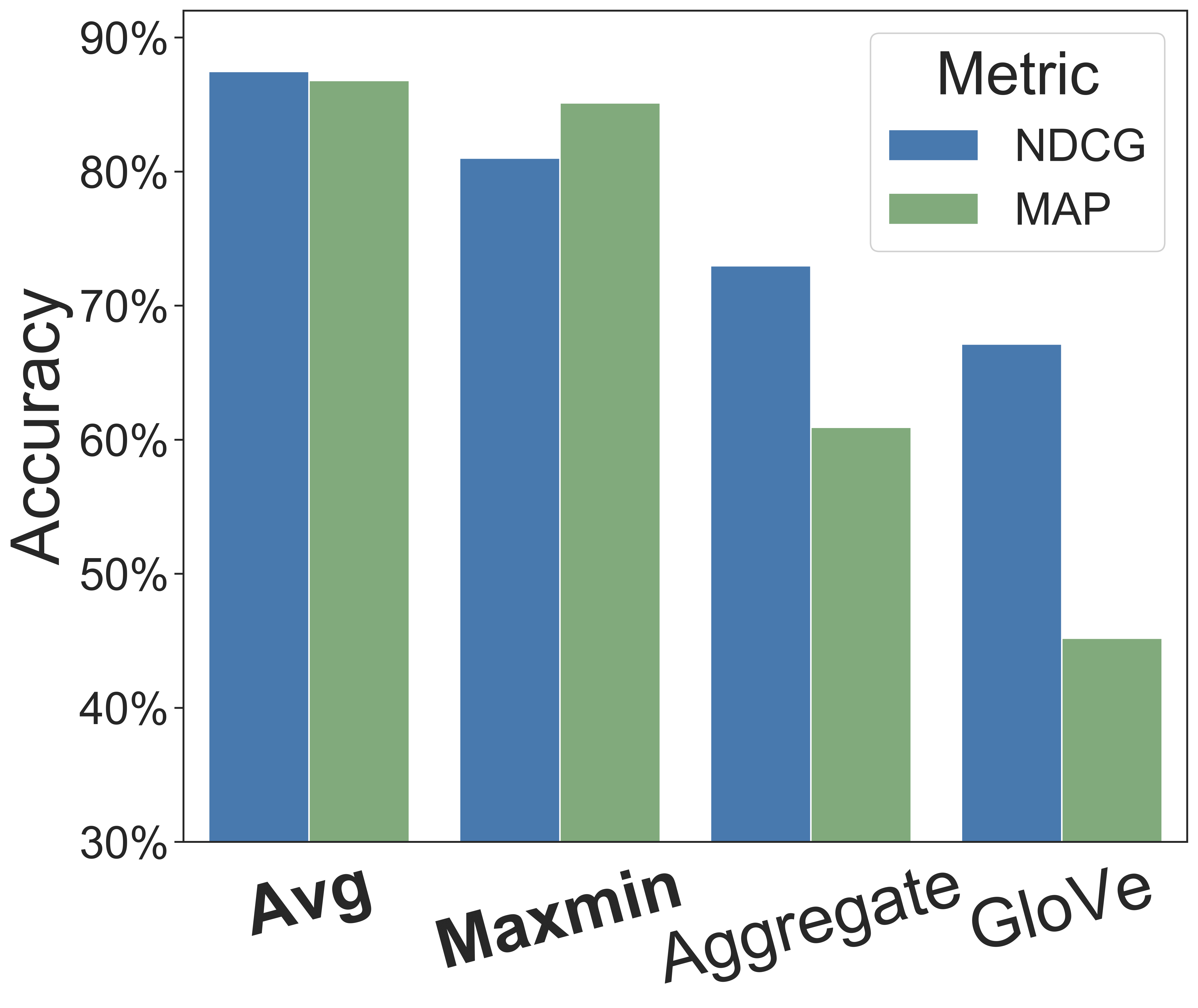}
    \caption{Results for search evaluation test case. Mean average precision (MAP) and Normalized Discounted  Cumulative Gain (NDCG) by method, averaged across queries. Methods in bold use our model. \label{fig:search_study}}
\end{figure}
See Figure \ref{fig:search_study} for results. We report Non Cummulative Discounted Gain (NDCG) and Mean Average Precision (MAP), two common metrics in information retrieval \cite{schutze2008introduction}.
Our FineGrained-AVG wins for both metrics, followed by FineGrained-MAXMIN. The baselines perform much worse, with the aggregate-vectors approach in \cite{hope2017accelerating} outperforming standard embedding-based retrieval with GloVe. Importantly, our approach achieves high MAP (85\% - 87\%) in {\it absolute terms}, in addition to a large relative improvement over the baselines (MAP of 40\%-60\%).

\xhdr{Qualitative Analysis} Table \ref{tab:searchres_ex} shows example results of FineGrained-AVG. For instance, a query for using light not for lighting results in laser-based billiard instructions. A query for using RFID not for locating or tracking results in an idea for an RFID-based lock, or RFIDs used at supermarket checkouts. To give an intuition for what might be driving our quantitative findings, we examine examples of retrieved results. 

For instance, with the query for using light for the non-standard purpose of cleaning, the top ranked result retrieved by FineGrained-AVG is a {\it UV Light Sterilizer}, with extracted purposes including {\it Sterilizes bacteria}, {\it Keep public and people healthy} and {\it Cleaner fresher air}, and the top result from FineGrained-MAXMIN is similarly a {\it Standalone bug zapper bulb} that uses {\it uv light/black light}. Conversely, the top result for both baselines (standard search and aggregate-vectors) is a {\it Toilet/Bathroom Light}, with ``a {sensor light that glows around your toilet}'' and ``{extra batteries if you lose electricity in the bathroom}''. It appears that both baselines were not able to accurately capture and disentangle purposes and mechanisms, despite the aggregate-vector being explicitly designed for that. 

More generally, it appears that the aggregate-vector approach squashes multiple purposes together by design into one soft, aggregate vector, which in this case includes concepts like {\it toilet} and {\it bathroom} that are somewhat topically related to cleaning. The aggregate approach had similar issues in the next product ideas it retrieved (e.g., {\it Switch that glows in the dark}, a {\it Dash Light to illuminate ash trays}).% Only the fifth result (out of the top five) was closer to being related to the query (a {\it LED lamp designed to look like a window} that {\it can keep air odorless} with {\it an electrostatic air purifier}), yet not precisely capturing the purpose of cleaning -- due to squashing together multiple concepts in one soft average (this product was also ranked as the top fourth result by the standard search baseline). %In contrast, the fifth result found by FineGrained-MAXMIN was a {\it die grinder} with a {\it light to see inside when cleaning / fixing root welds inside steel pipe}. 

% As another example, for the query of using RFID not for locating or tracking , the top result with both FineGrained-AVG and FineGrained-MAXMIN is a {\it walk through checkout scanner} that uses RFID, a product not captured by the two other baselines in their top five results. The first-ranked result found by the aggregate-vector baseline approach was a {\it customizable luggage system } with {\it RFID protection} (also the the second result retrieved by FineGrained-AVG) but it also retrieved products such as a {\it wifi enabled chip for kids and pets} that allows them to {\it go in or out without tripping the alarm}, and a {\it case with laser and bluetooth to connect to smart devices}, that are of weaker relatedness to RFID technology.

Overall, our results demonstrate that fine-grained purposes and mechanisms lead to better functional search expressivity than approaches based on distributional representations or coarse purpose-mechanism vectors.
%  outperforming effective \cite{arora2017embed,reimers-2019-sentence-bert}

%% file: 06_Hier.tex
%In the previous section we evaluated our model in terms of accuracy and utility of extracted chunks. We now turn to our main motivation, helping people with \textbf{creative problem-solving}. %Given a problem from a user (e.g., query to search engine), we examine whether we can help boost ideation by \textbf{suggesting analogies} -- new, structurally related problems. 

%{Exploring the design space} to find related problems is a critical subtask of innovation.
In this section we test the value of our novel representation for supporting users in exploring the design space for solving a given problem. We use our span-based representation to construct a corpus-wide graph of purpose/mechanism concepts. We demonstrate the utility of this approach in an ideation task, helping users identify useful inspirations in the form of problems that are related to their own.

Our goal is to help users ``break out'' of fixation on a certain domain, a well-known hindrance to innovation \cite{chanBestDesignIdeas2015,kittur2019scaling}. Doing so is challenging because it requires some level of \emph{abstraction}: being able to go beyond the details of a concrete problem to connect to a part of the design space that may look dissimilar on the surface, but has abstract similarity. Numerous studies in engineering and cognitive psychology have shown the benefits of problem abstractions for ideation \cite{linsey2012design, fu_expert_2013,fu_discovering_2013,yu2014searching,yu2014distributed,kittur2019scaling,goucher2019crowdsourcing}. However, these studies either involve non-scalable methods (relying on highly-structured annotations, or on crowd-sourcing) or simple, syntactical pattern-matching heuristics incapable of capturing deeper abstract relations. In  \cite{hope2017accelerating} (aggregate-vectors baseline from the previous section), crowdworkers were given a product description from the Quirky database, and asked to come up with ideas for products that solve the same problem in a different way. Aggregate vectors representing purpose and mechanism were used to find near-purpose, far-mechanism analogies. Thus, finding analogies relied on having a given mechanism to control for structural distance. 

Unlike \cite{hope2017accelerating}, in our setup we assume a more realistic scenario where we are given only a short problem description -- e.g., \textit{generating power for a phone, reminding someone to take medicine, folding laundry} -- and aim to find inspirational stimuli \cite{goucher2019crowdsourcing} in the “sweet spot” for creative ideation -- structurally related to the given problem, not too near yet also not too far \cite{fu_meaning_2013}. 

\xhdr{Functional Concept Graph} To address this challenge, we build a tool inspired by functional modeling \cite{hirtz2002functional}, which we call a \emph{Functional Concept Graph}. A functional model is, roughly put, a hierarchical ontology of functions and ways to achieve them, and is a key concept in engineering design. Such models are especially useful for innovation, allowing problem-solvers to break out of a fixed overly-concrete purpose or mechanism and move up and down the hierarchy. Despite their great potential, today's functional models are constructed manually, and thus do not scale.  While automatically inducing full abstraction hierarchies/ontologies of functional properties of real-world products remains a daunting challenge, in our approach we construct a rough approximation --- simple enough to extract automatically from noisy product texts, while still being useful for exploring the design space and suggesting inspirations to users. Specifically, in our approach Functional Concept Graphs consist of nodes corresponding to purposes or mechanisms, and edges reflect semantic relatedness that is not guaranteed to directly encode abstraction. We build this graph by observing fine-grained co-occurrences of concepts appearing together in products, using rule-mining to infer which concept is likely to be more general to (roughly) capture different levels of abstraction.

\begin{figure}
    \includegraphics[width=0.99\linewidth]{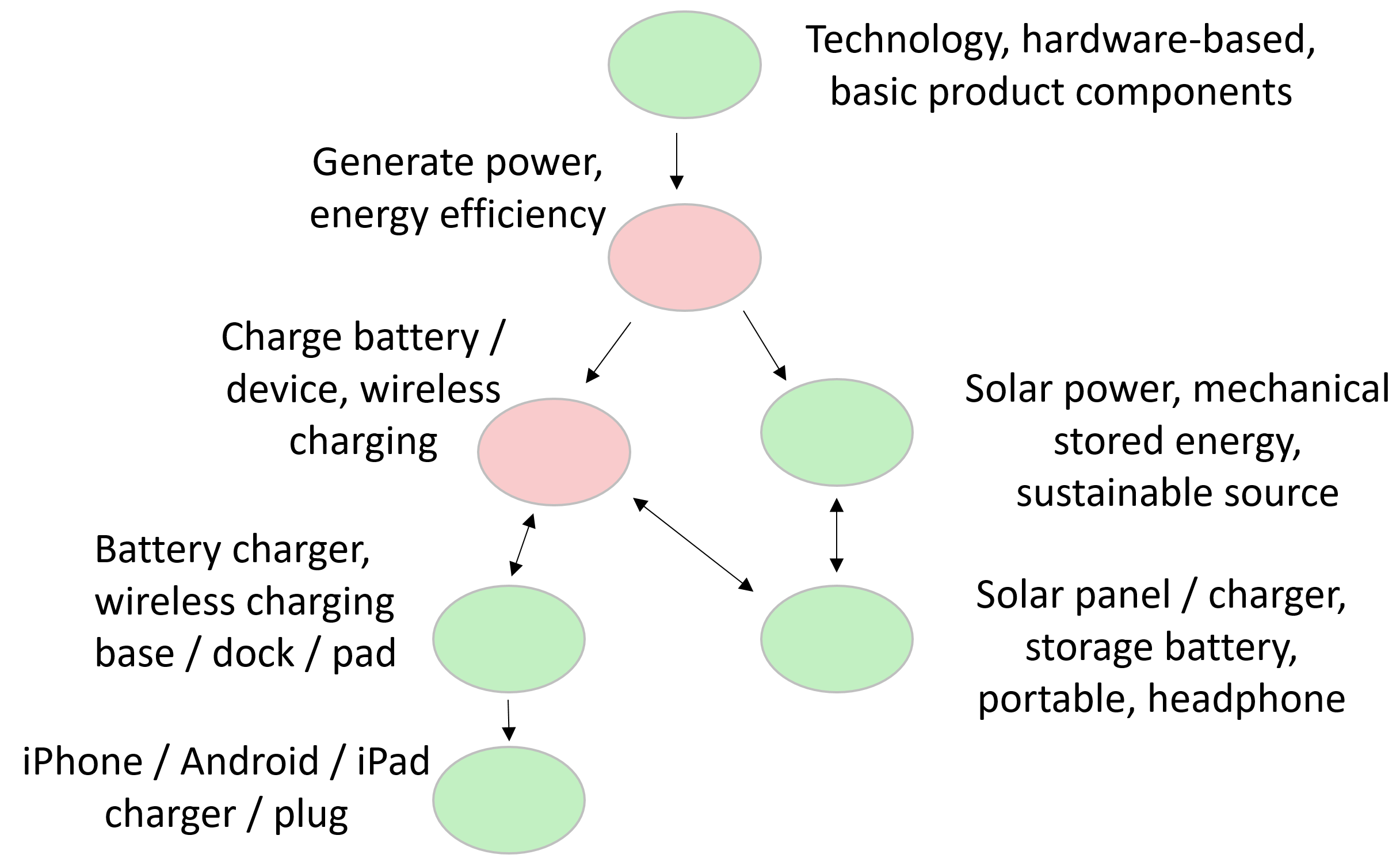}
    \caption{An example of our learned functional concept graph extracted from texts. Mechanism in green, purpose in pink. Titles are tags nearest to cluster centroids (redacted to fit).\label{fig:hier2}}
\end{figure}

For example, Figure \ref{fig:hier2} shows an actual subgraph from our automatically constructed functional concept graph related to electricity, power and charging. Products that mention certain purposes (e.g., ``charge your phone") will often mention other, structurally related problems that could be more general/abstract (e.g., ``generate power") or more specific (``wireless phone charging"), resulting in edges in our graph (only high-confidence edges are shown). 
 %Pink nodes correspond to purposes and green nodes to mechanisms. Only high-confidence edges are shown. %One can see a purpose node about generating power which has a sub-purpose of charging batteries and devices. There is a bidirectional edge between that cluster (charging batteries) and a mechanism cluster about chargers. This mechanism cluster (chargers) has a child cluster specific to Apple chargers. 
A designer could go from the problem of charging batteries to the more general problem of generating power, and from there to another branch  (e.g., solar power and mechanical stored energy), to get inspired by structurally related ideas. %In the experiment that follows, we focus only on purpose-purpose relations $\mathcal{C}_p \times \mathcal{C}_p$.

\remove{
In other words, we can use the graph to discover patterns in the form of \textit{products that solve problem} $p_i$\textit{, also often solve problems} $\mathcal{I}$, and suggest $\mathcal{I}$ as potential inspirations to be recommended \footnote{This also bears certain resemblance to collaborative filtering \cite{koren2015advances}, where recommendations are based on the pattern: \textit{people who buy item X, also often buy Y};  in our case, instead of people and items, we have ideas written by people, and the problems they solve.}. 
}

\subsection{Building a Functional Concept Graph}
\label{sec:fcg}

%\begin{figure}[t]
%    \centering
 We develop a method to infer this representation from {co-occurrence patterns} of the fine-grained spans of text. 
 Naively looking for co-occurrences of problems may yield inspirations  too near to the original $p_i$, as many frequently co-occurring purposes tend to be very similar, while we are interested discovering the more abstract relations. In addition, raw chunks of text extracted from our tagging model have countless variants that are not sufficiently abstract and are thus sparsely co-occurring. We thus design our approach to encourage abstract inspirations. As an overview of our approach before presenting the technical details, we take the following two steps:

\xhdr{I. Concept discretization} %First, we need to define the nodes in our graph. %We group purposes extracted from our corpus into more general, abstract concepts. 
Intuitively, nodes in our graph should correspond to groups of related spans (``charging'', ``charging the battery'', ``charging a laptop''). To achieve this, 
we take all purpose and mechanism spans $\Hat{\mathcal{P}}$,  $\Hat{\mathcal{M}}$ in the corpus, extracted using our GCN model, %(note that this could not be done with the previous work in \cite{hope2017accelerating} that extracted soft aggregate vectors). 
and cluster them (separately), using pre-trained vector representations. We refer to the %and consider cluster centers as our concepts 
clusters $\mathcal{C}_p, \mathcal{C}_m$ as \emph{concepts}.  %Original problems $p_i$ are also mapped to their corresponding clusters. 

%To further encourage abstraction, we give preference to clusters that not only co-occur with the original problem $p_i$, but who are also likely to be broader and more general than the cluster to which $p_i$ belongs. We do so with an intuitive \textbf{rule-mining} approach to find hierarchical relations between purpose clusters $\mathcal{C}_p$. 

\xhdr{II. Relations} %We next define our edges. %We use an intuitive \textbf{rule-mining} approach to find hierarchical relations between clusters. 
We employ rule-mining \cite{pasquier1999discovering} to discover a set of relations $\mathcal{R}$ between concepts (see §\ref{sec:ideationexp} for implementation details). Relations are \textit{Antecedent $\Longrightarrow$ Consequent}, with weights corresponding to rule confidence. To illustrate our intuition, suppose that when ``prevent head injury'' appears in a product description,  the conditional probability of ``safety'' appearing too is large (but not the other way around). In this case, we can (weakly) infer that preventing head injuries is a sub-purpose of ``safety''.

Indeed, manually observing the purpose-purpose edges, the one-directional relations captured are often \emph{sub-purpose}, and the bi-directional ones often encode \emph{abstract similarity}. Similarly, for mechanism concepts the one-directional relations are often \emph{part of} (``cell phone'' and ``battery''), and bi-directional are mechanisms that \emph{co-occur} often. For pairs of purpose and mechanism concepts, the relation is often \emph{functionality} (``charger'', ``charge''). Exploring more relations is left for future work.

%% file: 07_hier_eval.tex
\remove{There exist many approaches to learn hierarchies from texts, often relying on advanced topic models \cite{movshovitz2015kb,zhu2017unsupervised} that hierarchically cluster spans of texts into \textit{general} concepts and discover a (large) set of relations in various approaches. Our focus is on testing whether our model's extracted annotations are useful for generating abstract analogical inspirations. We employ a rule-mining approach based on the outputs of our model. 

For an intuitive example, if we see many patents mentioning ``seat belts'' in the mechanism and ``safety'' as a (sub)purpose, we can infer a connection between the two. Similarly, if we see that when ``prevent head injury'' appears in the purpose,  the conditional probability of ``safety'' is large (but not the other way around), we can infer that preventing head injuries is a sub-purpose of ``safety''. We use such observations in our approach. }

\remove{Our focus is on testing whether our model's extracted annotations are useful in building a commonsense hierarchy, and exploring its utility for generating abstract analogical inspirations; that is, we do not compete with the myriad generic ontology-construction methods available.}

\remove{\xhdr{Implementation and intrinsic evaluation} We represent our spans with GloVe word embeddings \cite{pennington2014glove}, normalized and averaged over tokens within a span. We cluster using K-Means, with $K=250$ selected automatically with an elbow-based criteria on silhouette scores. We then apply the FP-growth algorithm and select rules with confidence over a threshold.  

We recruited an engineer and a research scientist as raters. We constructed three sets of $20$ pairs each for purpose-purpose, mechanism-mechanism, and purpose-mechanism relations. The pairs were selected by ranking with respect to rule confidence. We asked judges to rate each pair as either good or bad, where a relation is considered good if it fits the \textit{Antecedent $\Longrightarrow$ consequent} expectation of the raters.

To ground our results, we compare to a simple baseline involving basic syntactic information \cite{fu2013discovering} that provides weak-but-reasonable purpose/mechanism signal: We extract noun-phrases as potential mechanism candidates and verbs as purpose candidates, and perform the same clustering and rule-mining steps. Results were randomized and judges were blind to condition. 

\xhdr{Results} Our model's accuracy was relatively high in absolute terms ($70\%$ for mechanism-mechanism and purpose-mechanism relations, and $80\%$ for purpose-purpose rules) and outperforms the syntax-based approach ($7\%$, $30\%$ and $20\%$, respectively), which often led to concepts too abstract, incoherent, or noisy. Judge agreement was substantial, with Cohen's kappa at $47\%$ (though we observe a moderate degree of mislabeling). 
}

\subsection{Study Design \& Implementation}
\label{sec:ideationexp}
%\xhdr{Study Design}
Next, we set out to test the utility of the functional concept graph in an ideation task.
In our setup participants are given problems (e.g., reminding people to take their medication in the morning) and are asked to think of creative solutions. Participants were also given a list of \emph{potential inspirations}, grouped into boxes, and were instructed to mark whether each was novel and helpful. They were also encouraged to explain the solution it inspired. 

\begin{figure}[t!]
    \centering
    \includegraphics[width=0.99\columnwidth]{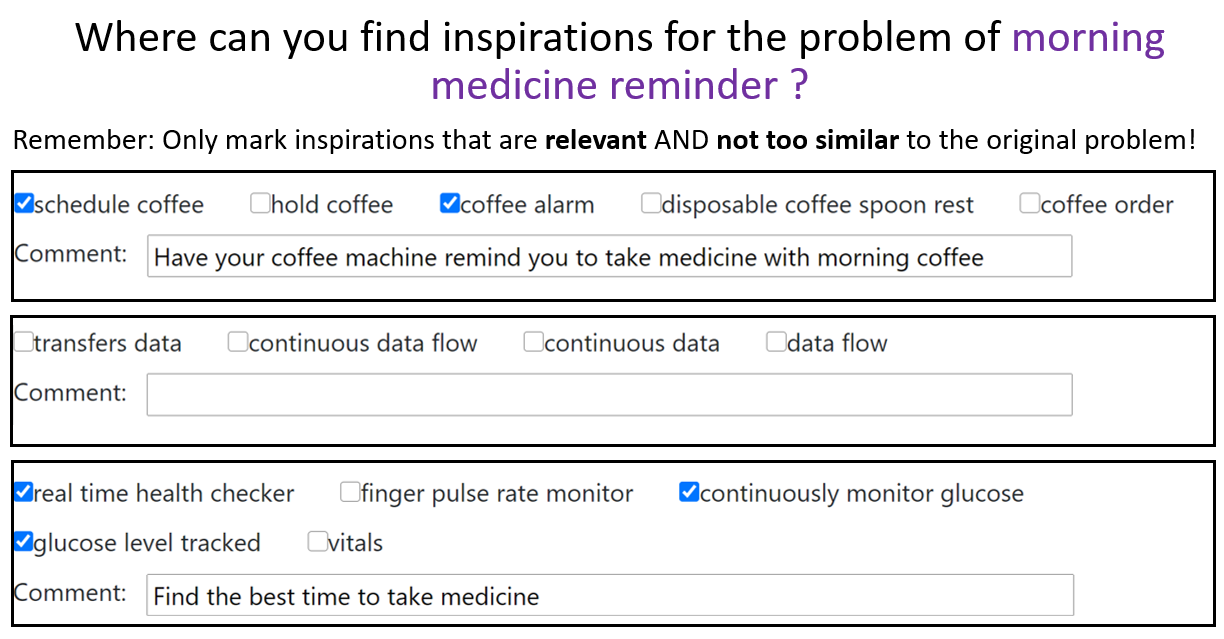}
    \caption{A snippet from our ideation interface for ``morning medicine reminder''. Users see inspirations grouped into boxes. Each box is supposed to represent a concept -- a cluster of related spans as found by our method or by the baselines (see §\ref{sec:fcg}). Users indicate which inspirations were useful, and what ideas they inspired. For example, seeing ``real time health checker'' inspired one user to suggest a monitoring device for finding the best time for reminding to take the medicine. \label{fig:ideation}} \end{figure}

Figure \ref{fig:ideation} shows our interface. In this example, seeing inspirations about health monitors caused one user to suggest monitoring the person to find the best time to remind them to take medicine; seeing inspirations about coffee caused them to suggest integrating medicine reminders into coffee machines. 

To create a set of {\bf seed problems}, a graduate student mapped between problems from WikiHow.com (a website of how-to guides) to purposes in our data. Using this source allowed us to collect real-world problems that are broadly familiar, with succinct and self-explanatory titles that do not require further reading to understand. The student was tasked with confirming that our Quirky dataset contains idea descriptions that mention these problems. For a given problem in WikiHow (\textit{how to remember to take medication}), they performed keyword search over $17K$ purpose spans gleaned by our model from Quirky, and found matching spans (\textit{morning medicine reminder}). We use those matching spans as our seed problem description given to users (purple text in Figure \ref{fig:ideation}). We collect $25$ problems this way. Table \ref{tab:inspir_ex} shows more examples, such as \textit{Tracking distance walked}, \textit{folding laundry} or \textit{sensing dryness level}.

{\bf Inspirations} are other purpose spans from our dataset (see Table \ref{tab:inspir_ex}), selected automatically using our approach or baseline approaches.

 %. For example, for the problem of \emph{morning medicine reminder}, one user was inspired by ``schedule coffee'' to suggest a programmable coffee machine with a timer, that can remind you to take your medicine while dispensing your coffee in the morning. In the following, we explain our approach toward finding inspirations. For each of the 25 problems, we generate inspirations using the following methods:

%\xhdr{Problem collection} 

%\subsubsection{Computing Inspirations} 
%Now that we have our concept graph, we can suggest inspirations to users.  
 % with our rule-mining graph, as well as with baselines. 

 %, enabling to find purposes that are related but not too similar to the original.  %When observing bidirectional relations, we take the direction with maximal confidence (corresponding to conditional likelihood).
%or the given problem $p_i$, we first map it to a cluster in $\mathcal{C}_p$ ($\mathcal{C}_m$), and then take its \textit{consequents} as inspirations.  
\xhdr{Our Method} 
For our approach, we construct a functional concept graph as in Section \ref{sec:fcg}. To cluster related  spans into concept nodes, we explore two common and powerful vector representations of spans
%, one based pre-trained word embeddings  and the other on a more recent language model representation tuned 
to capture semantic similarity:

%In more concrete detail, we experiment with the following approaches for selecting inspirations. We experiment with two span representations for building a functional concept graph:
{
\setdefaultleftmargin{0.5em}{0.5em}{}{}{}{}
\begin{compactitem}

    \item \textbf{GloVe} \cite{pennington2014glove} pre-trained word embeddings, averaged across tokens.
    \item \textbf{BERT-based} \cite{reimers-2019-sentence-bert} contextualized vectors that have been fine-tuned for semantic similarity tasks\footnote{We use RoBERTa-large-STS-SNLI, available at \url{github.com/UKPLab/sentence-transformers}.}.
\end{compactitem}
}

We cluster the spans using K-Means++ \footnote{$K=250$ selected automatically with elbow-based criteria on silhouette scores.}\cite{arthur2007k}. We then apply the Apriori algorithm\footnote{\url{http://www.borgelt.net/pyfim.html}.} to automatically mine association rules between clusters,
\cite{pasquier1999discovering} and use the confidence metric to select the top rules\footnote{We use the top $3$ rules in our experiment.}. To use the mined rules between purpose nodes (clusters) for selecting inspirations shown to users, we start from the purpose node corresponding to the given problem and take its \textit{consequents}; as explained earlier, this captures a weak signal of abstract similarity.

\begin{table*}
%\begin{footnotesize}
\begin{tabular}{c|c|c}
    \hline
    \textbf{Problem} & \textbf{Inspirations} & \textbf{Rater explanation} \\
    \hline
    \multirow{1}{*}{Track distance walked}& \parbox{0.4\columnwidth}{Protect children} & \parbox{1\columnwidth}{Get ideas from devices that keep track of children} \\
    \hline
    \multirow{1}{*}{Folding laundry}& \parbox{0.4\columnwidth}{Store toilet paper} & \parbox{1\columnwidth}{Roll laundry around a tube instead of folding} \\
    \hline
    \multirow{1}{*}{Dispense medicine} &\parbox{0.4\columnwidth}{Pet bowl that keeps ants away} &\parbox{1\columnwidth}{Based on pet bowls that can dispense food during the day}\\
    \hline
     \multirow{3}{*}{Sense dryness level} & 
     \parbox{0.4\columnwidth}{Voltage reading} & \parbox{1\columnwidth}{Use electric current to measure water level (safely)} \\\cline{2-3}
     &\parbox{0.4\columnwidth}{Waterproof} & \parbox{1\columnwidth}{Ideas from sensors in waterproof devices}\\\cline{2-3}
     &\parbox{0.4\columnwidth}{Temperature reading} & \\
    \hline 
    \multirow{3}{*}{Morning medicine reminder}& \parbox{0.4\columnwidth}{Schedule coffee, coffee alarm} & \parbox{1\columnwidth}{Alarm clock with coffee and medicine reminders} \\\cline{2-3}
     &\parbox{0.4\columnwidth}{Send vital data, real-time health checker} & \parbox{1\columnwidth}{ Health trackers to tell if medicine not taken, alert accordingly} \\\cline{2-3}
     &\parbox{0.4\columnwidth}{Heart rate monitoring, continuously monitor glucose} & \parbox{1\columnwidth}{Find the best time to take medicine} \\
     \hline
    
\end{tabular}
%\end{footnotesize}
\caption{Example inspirations and explanations given by human evaluators.}
\label{tab:inspir_ex}
\end{table*}

Some of these nodes contain tens of spans in them. Thus, we also explore two approaches to ``summarize'' each concept cluster with representative spans displayed to users -- one that attempts to summarize the cluster independently of the seed problem, and one that takes the seed problem into account: 

{\setdefaultleftmargin{0.5em}{0.5em}{}{}{}{}
\begin{compactitem}
    \item \xhdr{TextRank \cite{mihalcea2004textrank}} We construct a graph where nodes are the spans in a cluster and edges represent textual similarity. We run PageRank \cite{page1999pagerank} on this graph, selecting the top $K$ spans to present.
    \item \xhdr{Nearest spans} Following the findings in \cite{fu_meaning_2013}, %suggesting a ``sweet spot'' for distance from the design problem when choosing analogical stimuli, 
    select the top $K$ spans in $\mathcal{C}_p$ that are \textbf{nearest} to the query $p_i$. (For both approaches, we use $K=5$).
\end{compactitem}
}
\xhdr{Baselines}\footnote{We note that all methods and baselines include both single and multi-word spans of text as inspirations, ensuring users are blind to the condition.}
{ \setdefaultleftmargin{1em}{1em}{}{}{}{}
\begin{compactitem}

\item \xhdr{Purpose span similarity} Given a problem $p_i$, we find the $K=5$ nearest purpose spans of text in our corpus (out of $17K$ purposes).
We experiment with the same two vector representations used by our approach: GloVe and BERT. This method is similar to applying the methodology in \cite{hope2017accelerating} to our setting, where in our setting we are given only a problem $p_i$ and no mechanism $m_i$ is available to control for structural distance. While this approach relies on our model for extracting purpose spans, we consider it a baseline to study the added value of our hierarchy. 
  \item \xhdr{Linguistic abstraction} We use the WordNet \cite{miller1995wordnet} lexical database to extract hypernyms (for each token in $p_i$), in order to capture potential abstractions. WordNet is often used in similar fashion for design-by-analogy studies \cite{linsey2012design,gilon2018focab,goucher2019crowdsourcing}.
 \item \xhdr{Random concepts} Random inspirations are often considered as a baseline in ideation studies since diversity of examples is a known booster for creative ability \cite{hope2017accelerating}. For each task, we select a random cluster from $\mathcal{C}_p$ and display its TextRank summary.
\end{compactitem}
}

\xhdr{Study Participants} We recruit $10$ raters to evaluate the inspirations, via university mailing lists. $8$ raters were engineering graduate students, and the remaining two raters included a senior engineering professor and an architect. This cohort is intended to reflect a target user base of people interested in innovation and involved in creative inventive thinking as part of their work.

\xhdr{Rating Collection} In our study, each method generated $K=5$ spans (concept summaries), which are grouped and displayed together in a box (see Figure \ref{fig:ideation}). For each problem a rater views $8$ boxes in randomized order, to avoid bias.  Raters were instructed to mark inspirations they consider useful and relevant for solving a given problem, while being \textit{not about the same problem}. Raters were also encouraged to write comments, especially for non-trivial cases which they found of interest (see Table \ref{tab:inspir_ex}). In total, raters viewed $2584$ boxes, or $12920$ purpose descriptors.

\subsection{Results} 
\begin{figure}[t!]
    \centering
    \includegraphics[width=1\columnwidth]{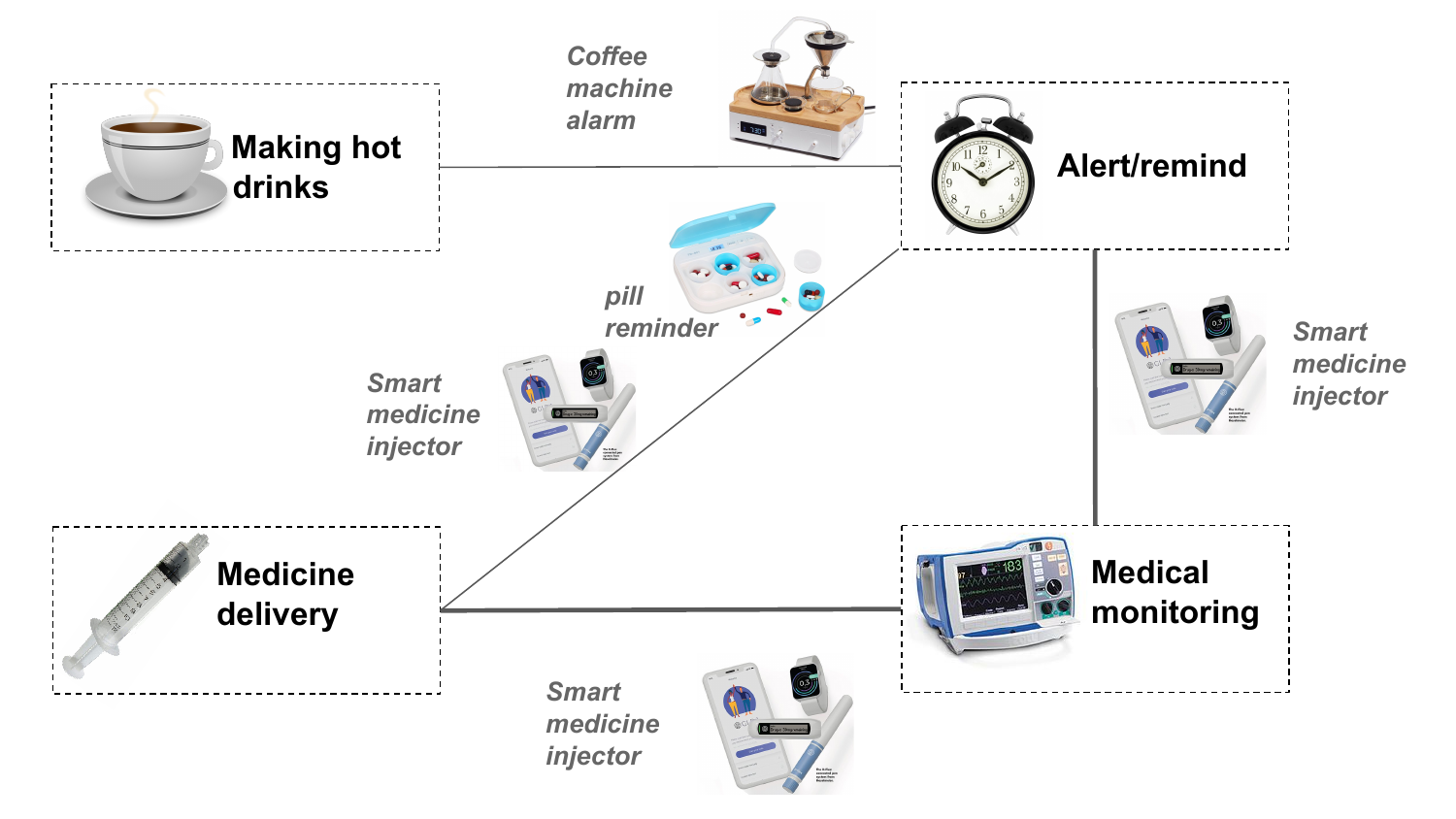}
    \caption{Excerpt from our Functional Concept Graph. Nodes represent concepts (clusters of purposes). To give intuition for how edges were created, they are annotated with example products containing spans from both concepts. All nodes and edges in this figure were automatically constructed and used to create the user-facing inspirations shown in Figure \ref{fig:ideation}. This figure provides a graphic illustration (not shown to users) explaining how the boxes in Figure \ref{fig:ideation} are generated, with node names provided here by us for readability. The problem of ``medicine morning reminder'' is mapped (via embedding) to the \textit{Alert/remind} concept cluster (as named by us), which is linked to the concepts of medical monitoring and making hot drinks through products such as ``smart medicine injector'' and ``coffee machine alarm'' (among others, not displayed in the figure). These links serve as the source for inspirations in our study, as seen in Figure \ref{fig:ideation}. \label{fig:medgraph}} \end{figure}

\xhdr{Analyzing inspirations} Table \ref{tab:inspir_ex} and Figure \ref{fig:ideation} show examples of problems, inspirations and user explanations from our study. For instance, users facing the ``morning medicine reminder'' problem were presented with nearby concepts in the Functional Concept Graph that included {health monitoring} and {coffee machines}. To explore why these concepts are connected in our graph and why they are potentially useful as inspirations, we make use of the direct interpretability of our approach. We examine the purpose co-occurrences from which the Functional Concept Graph was constructed. 

Figure \ref{fig:medgraph} shows the subgraph with concept nodes of \textit{Making hot drinks, alerting/reminding, health monitoring, medicine delivery}, and edges representing products in which two adjacent purposes were co-mentioned (e.g., a {``coffee machine alarm''} product that mentioned the purposes of \emph{making hot drinks} and \emph{alerting/reminding}, or a ``smart medicine injector" that mentioned both \emph{alerting/reminding} and \emph{medicine delivery}). This explains why the concepts are nearby in the graph, as there are multiple products in our dataset that share purposes from both concepts.

For example, a {``pill reminder''} product refers to the problem of forgetting to take medicine at prescribed times ({\it Sends notification if you forgot to take your AM or PM meds}), while a {``smart injector''} device administers medicine {\it on set time intervals}. At the same time, both of these products mention purposes of medicine delivery. When our graph construction algorithm observes enough similar co-occurrence patterns between the concepts of alerting and medicine delivery, across multiple products, an edge is added between the two in the graph. Similarly, an {``Alarm coffee maker''} product mentions the purposes of {\it time management} and {\it making coffee at a set time} as well as {\it alerting when the coffee is ready}, explaining how it emerges as a potential inspiration in our graph.

This type of linkage or overlap between an original problem space and inspiration problems helps get at a sweet-spot of innovation \cite{chan2011benefits} by finding ideas that are not too near and not too far from the original problem, helping users break out of fixation as discussed earlier in this section. Users in our study used these inspirations to come up with a tracker that alerts the user at the best time to take a medicine and a coffee machine reminding the user to take their medication with their morning coffee. Those creative directions demonstrate the utility of the Functional Concept Graph for exploring the design space.

\begin{figure*}
    \includegraphics[width=1\columnwidth]{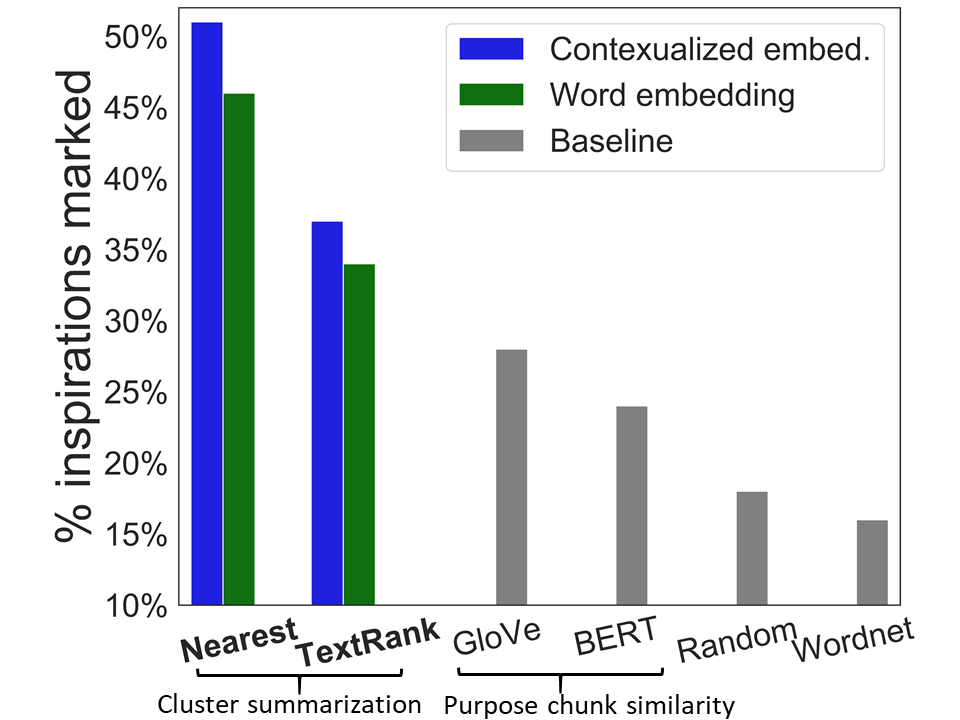}
    \includegraphics[width=1\columnwidth]{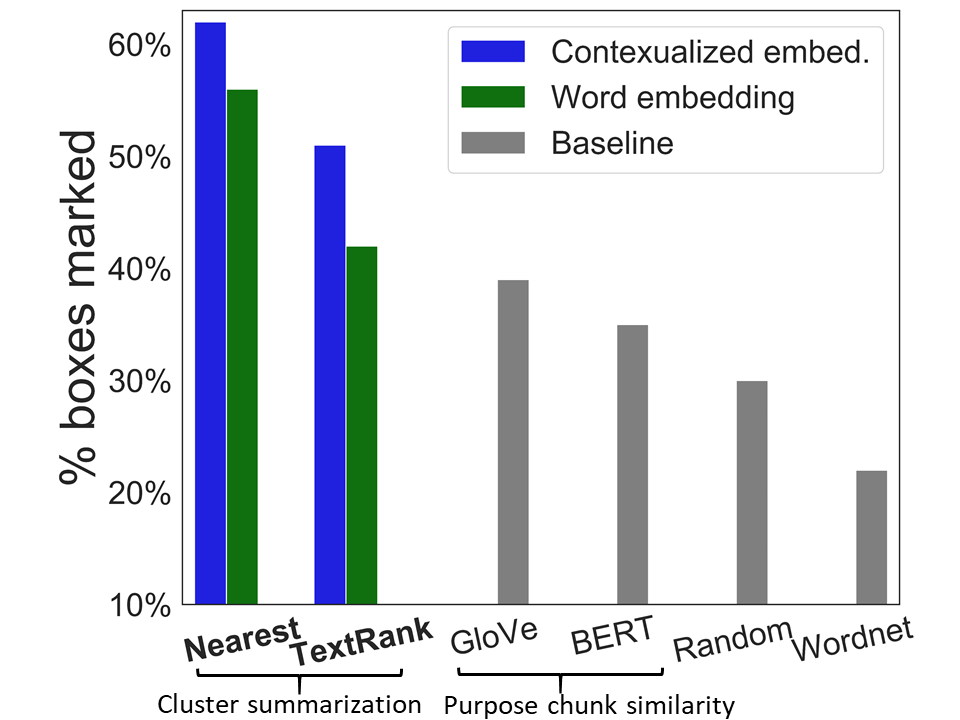}
    \caption{Inspiration user study results. Left: Proportion of inspirations selected by at least $2$ raters, per condition. Right: Proportion of boxes (clusters) with at least $2$ spans marked by $\geq 2$ raters. \label{fig:insp_study}}
\end{figure*}

\xhdr{Quantitative results} Figure \ref{fig:insp_study} shows the results of the user study. On the left, we show the proportion of inspirations (individual spans) selected by at least two raters, for each method. Our approach significantly outperforms all the baselines. The effect is particularly pronounced for the BERT-based approach, with $51\%$ of inspirations found useful, while the best baseline reaches less than $30\%$. Interestingly, for both BERT and GloVe representations, the Nearest-span summarization approach fares better, potentially due to striking a balance between being too far/near the initial problem $p_i$. 

Figure \ref{fig:insp_study} (right) shows the proportion of inspiration \textit{boxes} that got at least $2$ individual inspirations marked (by at least $2$ raters). This metric measures the effect of a box as one unit, as each box is meant to represent a coherent cluster. Our method is able to reach $62\%$, while the best baseline (GloVe search on purpose spans) yields only $39\%$. Again, the nearest-span summarization is prefered to TextRank. Importantly, for both individual inspiration spans and inspirations boxes,  $51\%$-$62\%$ are rated as useful -- high figures considering the challenging nature of the task.

%\dnote{send some clusters, preferably both glove and BERT of the same thing. also one cluster with both summarization methods}

 %Nevertheless, we can help inject useful semantic structure about product purposes and mechanisms into those models, which they are usually blind to.
%

 %As in rule-mining, if for instance a purpose concept $C_p \in  \mathcal{C}_p$ is an antecedent in a relation where a mechanism concept $C_m \in  \mathcal{C}_m $ is a consequent, we except (with high probability) to observe $C_p $ whenever we observe $C_m$ mentioned in a product text, but not necessarily vice versa. 

\remove{
{ \setdefaultleftmargin{2em}{3em}{}{}{}{}
\begin{compactitem}
\item \xhdr{Extraction} For each product $i$ in a corpus $\mathcal{D}$, extract purpose and mechanism sets $\mathcal{P}$ and $\mathcal{M}$. Build sets of all purpose spans $\Hat{\mathcal{P}}$ and all mechanism spans $\Hat{\mathcal{M}}$. 
\item \xhdr{Cluster into Concepts} We run K-means ($K=100$) over each of $\Hat{\mathcal{P}},\Hat{\mathcal{M}}$ separately, resulting in $K$ cluster centers for purpose and $K$ for mechanism. We consider these as our concepts $\mathcal{C}_p, \mathcal{C}_m$. 
\item \xhdr{Mine concept hierarchies} To discover a set of hierarchical relations $\mathcal{R}$ needed to complete an ontology, we employ rule-mining using the FP-growth algorithm  and selecting rules with confidence over a threshold. Thus, our $\mathcal{R}$ consists of \textit{Antecedent $\Longrightarrow$ consequent} relations between extracted purpose concepts $\mathcal{C}_p \times \mathcal{C}_p$, mechanism concepts $\mathcal{C}_m \times \mathcal{C}_m$, and cross-relations $\mathcal{C}_p \times \mathcal{C}_m$. As in rule-mining, if for instance a purpose concept $C_p \in  \mathcal{C}_p$ is an antecedent in a relation where a mechanism concept $C_m \in  \mathcal{C}_m $ is a consequent, we except (with high probability) to observe $C_p $ whenever we observe $C_m$ mentioned in a product text, but not necessarily vice versa. 
\end{compactitem}}
}

\remove{
\subsection{Evaluation}
\label{hier_eval_subsec}
}
\remove{ 
To evaluate the effect of our extractions, we compare to a baseline involving basic syntactic information, which \cite{fu2013discovering} showed provides weak-but-reasonable purpose/mechanism signal. Following them, we extract noun-phrases as potential mechanism candidates and verbs as purpose candidates and perform the same clustering and rule-mining steps as above.}

\remove{
\xhdr{Results} Importantly, we aim for broad applicability of our methods. We do not require end-users of the hierarchy to be experts in functional modeling (or even familiar with it) to be able to use it.}

\remove{
\begin{figure}
    \centering
    \includegraphics[width=0.7\columnwidth]{hier_res_plot.pdf}
    \caption{Hierarchy extraction human evaluation. Average positive rating by pairing  (purpose-mechanism, purpose-purpose, mechanism-mechanism). We substantially improve upon the baseline.\label{fig:hier_ustudy}}
\end{figure}
}

%% file: 08_Discussion.tex
%\dnote{speculate a little about how this new representation could be used for other types of creativity support, and what would be needed to move to true abstraction.  I.e., what could be its impact.}

In this paper we introduced a novel span-based representation of ideas in terms of their fine-grained purposes and mechanisms and used it to develop new tools for creative ideation. 
We trained a model to extract spans from a noisy, real-world corpus of products. We used this representation to support human creativity in two applications: expressive search for alternative, uncommon uses of products, and generating a graph to help problem-solvers explore the design space around their problem. In both ideation studies, we were able to achieve high accuracies, significantly outperform baselines and \emph{help boost user creativity}.
%by controlling structural distance and finding ideas in the “sweet spot” for creativity (not too near, not too far).

% In future work, we would like to further explore weak supervision approaches to augment annotation in noisy settings. Another direction is learning purposes and mechanisms in an end-to-end fashion. 
% Another exciting prospect is deploying our search engine publicly, allowing scientists, engineers and designers to perform rich queries, discover new similarities, and boost innovation with enhanced capabilities not possible with today's search.
%\subsection{Limitations}

\subsection{Limitations} 
While our results showed the promise of a functional aspect-based representation, and demonstrated potentially feasible technical approaches for extracting this representation from unstructured text, the approach has several limitations.

\xhdr{Challenging Annotation Task for Crowds} First, the annotation task proved to be somewhat difficult for crowdworkers, and the outputs were noisy. One direction for future work would be to explore weak supervision approaches to augment annotation. One issue that might exacerbate the problem is that sometimes the boundary between purpose and mechanism is fuzzy, and it is genuinely difficult to tell how to annotate the span.

\xhdr{Limited Functional Schema} In a similar vein, it might be interesting to explore more expressive schemas, containing elements other than just purpose and mechanism (similar to \cite{chan2018solvent}).
One particularly useful element to add might be context/constraints (e.g., nanoscale), to restrict the search space to feasible solutions.   

\xhdr{Surface Form Abstraction} Another limitation of our approach is that our functional aspects (and resulting embeddings) remain quite closely anchored to the original texts. This limits their ability to be used to match across domains, to make connections such as inspiring new optimization approaches by analogy to "heating/cooling" schedules in metallurgy. Achieving abstraction to match across distant domains without burdening the user with a combinatorial explosion of noisy matches remains an open problem. We wonder if abstracting key objects or entities in a purpose functional aspect --- such as a more automated approach to replacing objects with their "commonsense" properties --- might be more feasible than attempting to abstract from an entire product description or abstract, given that the chunk is already a rich signal of the product's functional meaning.

\subsection{Future Work and Broader Implications}
%\dnote{richer semantics of edges, more experiments with people...}

Moving to future directions, we are excited about the potential of functional aspects to lead to advances in the interpretability of content-based recommender systems in these complex domains. Keywords are inherently interpretable, but are limited in their capacity to support crossing knowledge boundaries; and until now, embedding-based approaches (e.g., \cite{hope2017accelerating}) have not always led to interpretable justifications for matches. Functional aspects could provide the basis of not just more powerful search operators, but also more interpretable results and feedback loops.

\xhdr{Deeper Functional Graph Exploration} A key component for the above might be expanding on our use of functional graphs, built from the extracted functional aspects. In our experiments, we used our functional concept graph to retrieve inspirations from "around" the problem. But what would it take to be able to explore this graph? Could we identify and optimize for latent coordinates in the functional space, moving "up" and "down abstraction levels, or "across" sibling nodes in a functional graph? Taking inspiration from network perspectives on ideation \cite{caiExtendedLinkographyDistance2010, sosaAccretionTheoryIdeation2019,goncalvesLifeCycleCreative2021}, could we retrieve interesting "lineages" of ideas, or compute the potential inspiration value of functional aspects based on network connectivity metrics? Could we combine these content-based functional aspects with measures of use (e.g., citations), to enrich approaches that combine content- and social-based signals, such as literature-based discovery \cite{swansonUndiscoveredPublicKnowledge1996,smalheiserRediscoveringDonSwanson2017}? 

\xhdr{Identifying Overlap and Gaps Across Fields} These approaches rely on identifying interesting overlaps in concepts that simultaneously coincide with disjunctions in citations, as signals of potentially impactful "undiscovered public knowledge": a persistent challenge is how to define "concepts" - keyword or unstructured text approaches can lead to combinatorial explosions of noise to sift through, and controlled vocabulary (e.g., MEDLine) can help increase the signal to noise ratio, but are only available in specialized circumstances \cite{sebastianEmergingApproachesLiteraturebased2017}. Functional aspects might be a useful bridge between unstructured text and controlled vocabularies for identifying points of overlap and disjunction between different fields, accelerating the discovery of gems hidden in plain sight.

\xhdr{Functional aspects for Collaborative Ideation} Future work could also explore new interactions in collaborative and crowd innovation that might be enabled by the ability to quickly extract functional aspects in idea corpora. The source of our primary dataset here, Quirky, was actually a crowd innovation platform. HCI research on these platforms have begun to emphasize moving away from mere "selection" of best ideas from large samples of ideas, towards supporting generative collaboration over ideas .  Open problems include synthesizing major themes in large-scale corpora of user-generated ideas and identifying gaps in the idea space \cite{mahyarCivicDataDeluge2019,chanComparingDifferentSensemaking2016,siangliulueCollaborativeIdeationScale2015a}, as well as supporting intelligent matching and structuring ways for crowd innovators to collaborate and build on each others' ideas \cite{malonePuttingPiecesBack2017,chanImprovingCrowdInnovation2016a} between crowd innovators. We are excited about the potential for functional aspects to assist with these functions, as a complement to other approaches like crowd-powered synthesis \cite{siangliulue_ideahound:_2016,chiltonCascadeCrowdsourcingTaxonomy2013,girottoEffectPeripheralMicrotasks2017,andreCrowdSynthesisExtracting2014}. Here, too, the potential for functional aspects to be highly interpretable could power novel explorations of mixed-initiative systems for augmenting collaborative ideation at scale \cite{siangliulue_ideahound:_2016,mackeprangDiscoveringSweetSpot2019}.

\xhdr{Mapping of Design Spaces} Beyond supporting richer search for creative inspiration, a data-driven approach to extracting functional aspects and learning relationships between the aspects could power much more expansive approaches to mapping out design spaces for entire domains or problem areas, identifying key subproblems and constraints and novel paths through the design space. Mapping approaches like this, such as technological roadmapping \cite{boydenArchitectingDiscoveryModel2019}, have already shown significant promise for reinvigorating research and development in real-world applications such as neural recording \cite{marblestonePhysicalPrinciplesScalable2013}. However, these mapping exercises are still highly manual and labor-intensive processes; computational support for such tasks could have transformative impacts on innovation.

\remove{
Functional models are increasingly important in engineering, where they can guide problem decomposition and abstraction. Despite their importance, today's functional models are mostly hand-crafted and therefore small. 

In this paper we tackled the challenge of approximating functional modeling at scale for large design repositories.
We employ a sequence tagging model to extract multiple purposes and mechanisms from product descriptions, and use them to learn functional models of ideas at scale. We employ a graph convolutional network (GCN) architecture fusing both semantic and syntactic information, on challenging real-world datasets. 
We show that our extracted purposes and mechanisms are useful for (1) enabling expressive queries based on functional similarity,
and (2) building commonsense ontologies to map landscapes of ideas, taking a step towards building hierarchical functional models to enable problem abstraction and reasoning.

In future work, we would like to explore weak supervision approaches to augment annotation, as well as approaches for handling noise and partial annotations. Another direction is learning purposes and mechanisms with a hierarchy in an end-to-end fashion. 
%
%We are interested in exploring uses for our engine in diverse areas across science, engineering, and product design. 
One exciting prospect is deploying our search engine publicly, allowing scientists, engineers and designers to perform rich queries, discover new interrelations between ideas, and boost innovation with enhanced abstraction and reasoning capabilities not possible with today's search. }

%\dnote{ Describing the user study tasks more concretely (including a better fig 6 or remove fig 6 and describe in text)
%- Showing how the functional graph actually supports abstraction (e.g., in one of the examples in table 3).}

%\dnote{We note that the aggregate baseline \cite{hope2017accelerating} directly learns mechanism and purpose vectors, while in FineGrained our graph neural network is only used to extract spans which are embedded with pre-trained GloVe vectors, to test the added value in extracting finer-grained chunks. In future work, we would like to explore extracting abstract structures using our graph representations.}

%% file: 09_Supp.tex
\subsection{Model Details} \label{subsec:app_modeldetails}
\begin{figure}
    \centering
    \includegraphics{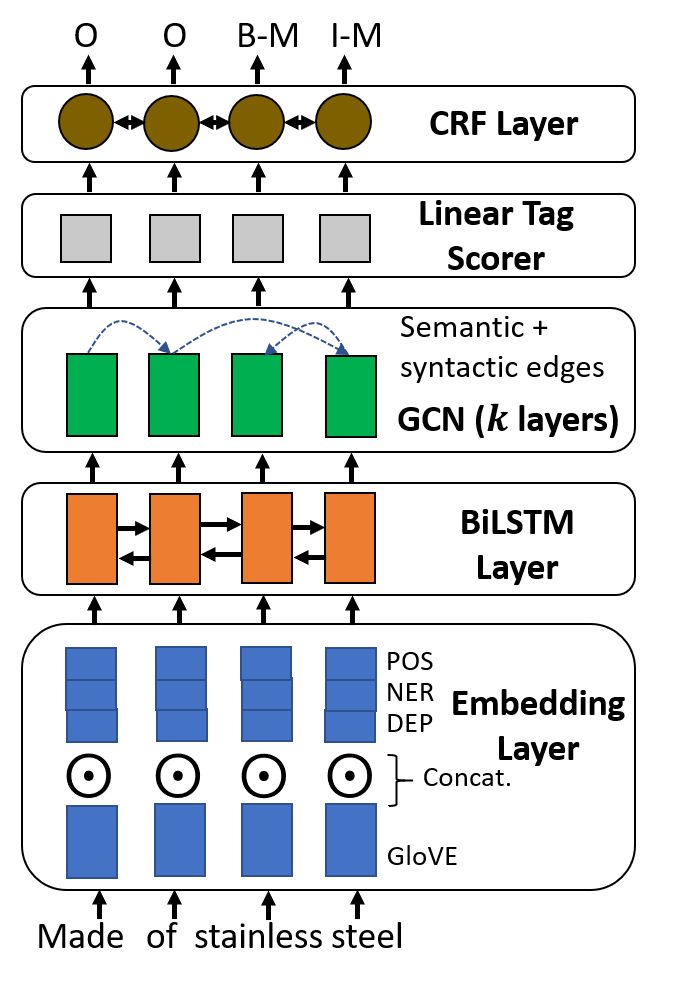}
    \caption{Schema of our GCN model.}
    \label{fig:modelschema}
\end{figure}

{
\setdefaultleftmargin{0.5em}{0.5em}{}{}{}{}
\begin{compactitem}
\item \xhdr{BiLSTM-CRF} A BiLSTM-CRF \cite{huang2015bidirectional} neural network, a common baseline approach for NER tasks, enriched with semantic and syntactic input embeddings known to often boost performance \cite{zhang2018graph}. We first pass the input sentence $\mathbf{x} = ({x}_1,{x}_2,\ldots,{x}_T)$ through an embedding module resulting in $\textbf{{v}}_{1:T}$, $\textbf{v}_{i}\in \Rbb^{d_e}$, where $d_e$ is the embedded space dimension. We adopt the ``multi-channel'' strategy as in \cite{zhang2018graph}, concatenating input word embeddings (pretrained GloVe vectors \cite{pennington2014glove}) with part-of-speech (POS) and NER embeddings. We additionally add an embedding corresponding to the incoming dependency relation. The sequence of token embeddings is then processed with a BiLSTM layer to obtain contextualized word representations $\textbf{h}_{1:T}^{\left(0\right)}$,  $\textbf{h}_{i}\in \Rbb^{d_h}$, where $d_h$ is the hidden state dimension. The outputs are fed into a linear layer $f$ to obtain per-word tag scores $f\left(\textbf{h}_{1}^{\left(L\right)}\right),f\left(\textbf{h}_{2}^{\left(L\right)}\right),...,f\left(\textbf{h}_{T}^{\left(L\right)}\right)$. These are used as inputs to a conditional random field (CRF) model which maximizes the tag sequence log likelihood under a pairwise transition model between adjacent tags \cite{akbik2018contextual}.

\item \xhdr{Pooled Flair} A pre-trained language model \cite{akbik2019pooled} based on contextualized string embeddings, recently shown to outperform powerful approaches such as BERT \cite{devlin2019bert} in NER and POS tagging tasks and achieve state-of-art results. Flair \footnote{\url{https://github.com/flairNLP/flair}} uses a character-based language model pre-trained over large corpora, combined with a memory mechanism that dynamically aggregates embeddings of each unique string encountered during training and a pooling operation to distill a global word representation. We follow \cite{akbik2019pooled} and concatenate pre-trained GloVe vectors to token embeddings, add a CRF decoder, and freeze the language-model weights rather than fine-tune them \cite{devlin2019bert,Peters2019ToTO}.
%\end{compactitem}
%}
%{
%\setdefaultleftmargin{0.5em}{0.5em}{}{}{}{}
%\begin{compactitem}
\item\xhdr{GCN} 
We also explore a model-enrichment approach with syntactic relational inputs. We employ a graph convolutional network (GCN) \cite{Kipf2016} over dependency-parse edges \cite{zhang2018graph}. GCNs are known to be useful for propagating relational information and utilizing syntactic cues \cite{zhang2018graph,Marcheggiani2017}. The linguistic cues are of special relevance and interest to us, as they are known to exist for purpose/mechanism mentions in texts \cite{fu_discovering_2013}.

We used a GCN with same token embeddings as in the BiLSTM-CRF baseline, with a BiLSTM layer for sequential context and a CRF decoder. For the graph fed into the GCN, we use a pre-computed syntactic edges with dependency parsing: For sentence $\textbf{{x}}_{1:T}$, we convert its dependency tree to $\textbf{A}^{syn}$ where $\textbf{A}_{ij}^{syn}=1$ for any two tokens $x_{i}, x_{j}$ connected by a dependency edge. We also add self-loops $\textbf{A}^{self}=I$ (to propagate from $\textbf{h}_{i}^{\left(l-1\right)}$ to $\textbf{h}_{i}^{\left(l\right)}$ \citep{zhang2018graph}). Following \cite{zhang2018graph}, we normalize activations to reduce bias toward high-degree nodes. For an $L$-layer GCN, denoting $\textbf{h}_{i}^{\left(l\right)}\in \Rbb^{d_h}$ to be the $l$-th layer output node, the GCN operation can be written as 
\[
h_{i}^{\left(l\right)}=\sigma\left(\sum_{r\in\mathcal{R}}\left[\sum_{j=1}^{n}\mathbf{A}_{ij}^{r}\mathbf{W}_{r}^{\left(l\right)}h_{j}^{\left(l-1\right)}/d^r_{i}+\textbf{b}_{r}^{\left(l\right)}\right]\right)
\]
 where  $\R=\left\{ \text{syn},\text{self}\right\}$, $\sigma$ is the ReLU activation function, $\mathbf{W}_{r}^{\left(l\right)}$ is a linear transformation, $\textbf{b}_{r}^{\left(l\right)}$ is a bias term and $d^r_{i}=\sum_{j=1}^{T}\mathbf{A}^{r}_{ij}$ is the degree of token $i$ w.r.t $r$. In the GCN architecture, $L$ layers correspond to propagating information across $L$-order neighborhoods.
We set the contextualized word vectors $\textbf{h}_{1:T}^{\left(0\right)}$ to be the input to the GCN, and use $\mathbf{h}_{1:T}^{\left(L\right)}$ as the output word representations. Figure \ref{fig:modelschema} illustrates the GCN model architecture. Similarly to \cite{Marcheggiani2017}, we do not model edge directions or dependency types in the GCN layers, to avoid over-parameterization in our data-scarce setting. We also attempted \textbf{edge-wise gating} \cite{Marcheggiani2017} to mitigate noise propagation but did not see improvements, similarly to \cite{zhang2018graph}.
\end{compactitem}
}

In our experiments, we followed standard GCN training procedures. Specifically, we base our model on the experimental setup detailed in \cite{zhang2018graph} (see also the authors' code which we adapt for our architecture, at \url{https://github.com/qipeng/gcn-over-pruned-trees}). We pre-process the data using the spaCy (\url{https://spacy.io}) package for tokenization, dependency parsing, and POS/NER-tagging. We use pretrained GloVE embeddings of dimension $300$, and NER, POS and dependency relation embeddings of size $30$ each, giving a total embedding dimension $d_e=390$. The bi-directional LSTM and GCN layers' hidden dimension is ${d_h}=200$, with $1$ hidden layer for the LSTM. 
We find that the setting of $2$ hidden layers works best for the GCNs. 
We also tried training with edge label information based on syntactic relations, but found this hurts performance. The training itself was carried out using SGD with gradient clipping (cutoff $5$) for $100$ epochs, selecting the best model on the development set. 

For the Pooled-Flair approach \cite{akbik2019pooled}, we use the FLAIR framework \cite{akbik2019flair}, with the  settings obtaining SOTA results for CONLL-2003 as in \cite{akbik2019pooled} (see \url{https://github.com/flairNLP/flair/blob/master/resources/docs/EXPERIMENTS.md}). We also experiment with non-pooled embeddings and obtain similar results. We experiment with initial learning rate and batch size settings described in \cite{akbik2019pooled}, finding $0.1$ and $32$ to work best, respectively. 

